\documentclass{article}
\usepackage[style=authoryear, sorting=ynt, sortcites=true, backend=biber, natbib=true, maxcitenames=2]{biblatex}
\DeclareDelimFormat[parencite]{nameyeardelim}{\addspace}
\usepackage[explicit]{titlesec}
\usepackage[title]{appendix}
\usepackage{booktabs}
\usepackage{amssymb, amsmath, bm}
\usepackage{subcaption, graphicx}
\usepackage{mathtools}
\usepackage[utf8]{inputenc}
\usepackage{subfiles}
\usepackage[letterpaper, margin=1in]{geometry}
\usepackage{xcolor}
\usepackage[colorlinks=true, linkcolor=black]{hyperref}
\usepackage{float}
\usepackage{setspace}
\usepackage{lineno}
\usepackage[british]{babel}
\usepackage{longtable}
\usepackage{rotating}

\usepackage{xr}

\newcommand{\RR}{\color{black}}
\newcommand{\RRR}{\color{black}}

\newcommand{\beq}{\begin{equation}}
\newcommand{\eeq}{\end{equation}}
\newcommand{\beqsub}{\begin{subequations}}
\newcommand{\eeqsub}{\end{subequations}}
\newcommand{\beqa}{\begin{eqnarray}}
\newcommand{\eeqa}{\end{eqnarray}}
\newcommand{\bize}{\begin{itemize}}
\newcommand{\eize}{\end{itemize}}

\begin{document}


\title{Modelling road mortality risks to persistence to a Western Toad ({\it Anaxyrus boreas}) population in British Columbia}
\author{Marguerite H. Mahr\thanks{EcoMosaic Environmental Consulting} \and Noah D. Marshall\thanks{Mathematics Department, McGill University} \and Jessa Marley\thanks{CMPS Department (Mathematics), University of British Columbia Okanagan} \and Sarah K. Wyse\thanks{Mathematics Department, Okanagan College} \and Wayne P. McCrory\thanks{McCrory Wildlife Services Ltd.} \and Rebecca C. Tyson\thanks{CMPS Department (Mathematics), University of British Columbia Okanagan}}

\maketitle

\begin{abstract}
Road mortality may be a significant factor in the global decline of amphibian populations, yet rigorous assessments of its effect on long-term population persistence are lacking.  
 Here, we investigate population persistence through a field study and mathematical model of a western toad ({\textit{Anaxyrus Boreas}} {\RR(Baird and Girard, 1852)}) population within a highway corridor in the Selkirk Mountains of British Columbia.
 The analysis shows traffic levels strongly correlate with toad mortality, with each additional vehicle causing a 3.1\% $\pm$ 1.3\% ($p=0.020$) increase in toad deaths.
 Although the current risk of the population becoming threatened or endangered is low, it rises to 50\% if baseline road mortality increases from 10\% to 30\%.  Gravid female mortality is higher than the baseline mortality and can increase the probability of endangerment by nearly two-fold at higher baseline mortality levels.
 We make the case that a small increase in vehicle traffic resulting from future development and recreational pressures
 could destabilize this apparently healthy toad population.  The high sensitivity to traffic levels and rapid transition from healthy to endangered raises concerns for similar populations worldwide.  Compensatory structures such as amphibian underpasses (toad tunnels) should be given high priority. 
\end{abstract}


\section{Introduction}
\label{sec:intro}

Amphibians, a key indicator of changing environmental conditions, continue to experience major population declines
worldwide \citep{GlobalDeclineAmphibians, habitatSplitGlobalDecline}. Particularly salient threats 
include climate change, habitat loss and fragmentation, forestry practices, and road mortality \citep{blaustein:1990, CauseAmphibianDecline, habitatSplitGlobalDecline, MPforWT:2014}. Amphibians are particularly vulnerable to road-related impacts because of their need to move between different habitats for their life history requirements \citep{bouchard:2009}.  In particular, many amphibians reside on land for most of the year, but migrate to and from aquatic habitat to breed.  As roads are often placed along water bodies,  they create a barrier to movement to and from the breeding areas, and put the migrating amphibians at risk of vehicle-induced mortality.  Understanding and mitigating the effects of road mortality on species persistence are important priorities in any species conservation effort \citep{taylor:2010, fahrig:2009}.

Amphibians are particularly vulnerable to road mortality \citep{fahrig:1995, glista:2008, chyn:2024}, and documenting road-amphibian interactions has been highlighted as an important activity in British Columbia \citep{BCMOE:2020}. Heightened understanding of the short- and long-term consequences of road mortality has led to the implementation of a variety of mitigation measures to increase amphibian safety and survivability.  These include underpasses, directional fencing, volunteer removal efforts, and public education \citep{BCMOE:2020}.  Models are an important tool in determining the long-term effects of road mortality \citep{barbosa:2020, jaeger:2005}, but relatively
little mechanistic modelling work has been done to quantify the effect of road mortality on amphibian population persistence
\citep{petrovan:2019, winton:2020} in spite of its documented importance \citep{hels:2001, jaeger:2005, ochs:2024}.  

We focus on a western toad ({\it Anaxyrus boreas} {\RR(Baird and Girard, 1852)}) population located at Fish and Bear Lakes in the Central Selkirk Mountains in the southern interior of British Columbia.  {\RR In North America, the western toad is widely distributed west of the Rocky Mountains \citep{natureserveexplorer:2022}.  Its conservation status varies from Critically Imperiled (S1) to Apparently Secure (S4) \citep{natureserveexplorer:2022}, reflecting regional disparities in population health and threats.  The Committee on the Status of Endangered Wildlife in Canada (COSEWIC) categorizes the western toad as a species of Special Concern \citep{cosewic:2012}.
%
%
Given that amphibians worldwide are facing steep and unforseen declines \citep{blaustein:1990}, there is urgency to conserving populations while they remain apparently healthy \citep{chiacchio:2022}.  The situation for western toads in southern British Columbia is a good example of this larger issue, as population declines and extirpations persist in this region \citep{BCMOE:2020}.}

In addition to being a listed species for which population studies are needed, western toads are an ideal species to study because the adults have high fidelity to their breeding site, and migrate annually and in sufficient numbers for relatively easy detection and observation.  {\RR Furthermore, our combined expertise allows us to use both field data and a mathematical model to assess extirpation risk.}  We analyse data from a six-year field study and develop a deterministic mathematical model of the population dynamics.  With these tools we estimate the potential impact of vehicle-induced mortality on population persistence.   

Most existing amphibian population models are either matrix or individual-based models with heavy data requirements \citep{petrovan:2019, jolivet:2008}, though a few  stage-structured discrete time difference equation models do exist (e.g.~\citet{jones:2017}). We present a stage-structured ordinary differential equation (ODE) model 
with migration events represented as impulses.
While our approach can be applied to any amphibian population whose main and breeding habitats are separated by a road, we focus here on the Fish and Bear Lakes western toad population as a case study.  {\RR We are particularly interested in assessing the extirpation potential for 
amphibian populations
under current and future traffic scenarios, when these populations need to cross a busy road as part of their life cycle.}
We use the model to estimate the level of road mortality at which extirpation is likely, and show that the impact of increased traffic depends critically on the mortality risk of gravid (i.e., egg-bearing) females.  We can then infer the potential benefit of reducing road mortality through mitigation measures such as amphibian underpasses with diversion fencing.

{\RR This paper presents both field and modelling studies and is structured to describe both.  For the field research component, we close this introduction with detailed descriptions of the study area and western toad life cycle (Sections~\ref{sec:studysite} and~\ref{sec:life+migration}).  In the Methods (Section~\ref{sec:methods}) we describe first the field study and then the mathematical model and its analysis.  The Results (Section~\ref{sec:results} cover the statistical analysis of the field data and then the mathematical model results.  We close (Section~\ref{sec:discussion}) with a discussion of road mortality effects on this population and potential implications for other amphibian populations.}

\subsection{Study Area}
\label{sec:studysite}

\begin{figure}
    \centering
    \includegraphics[width=\textwidth]{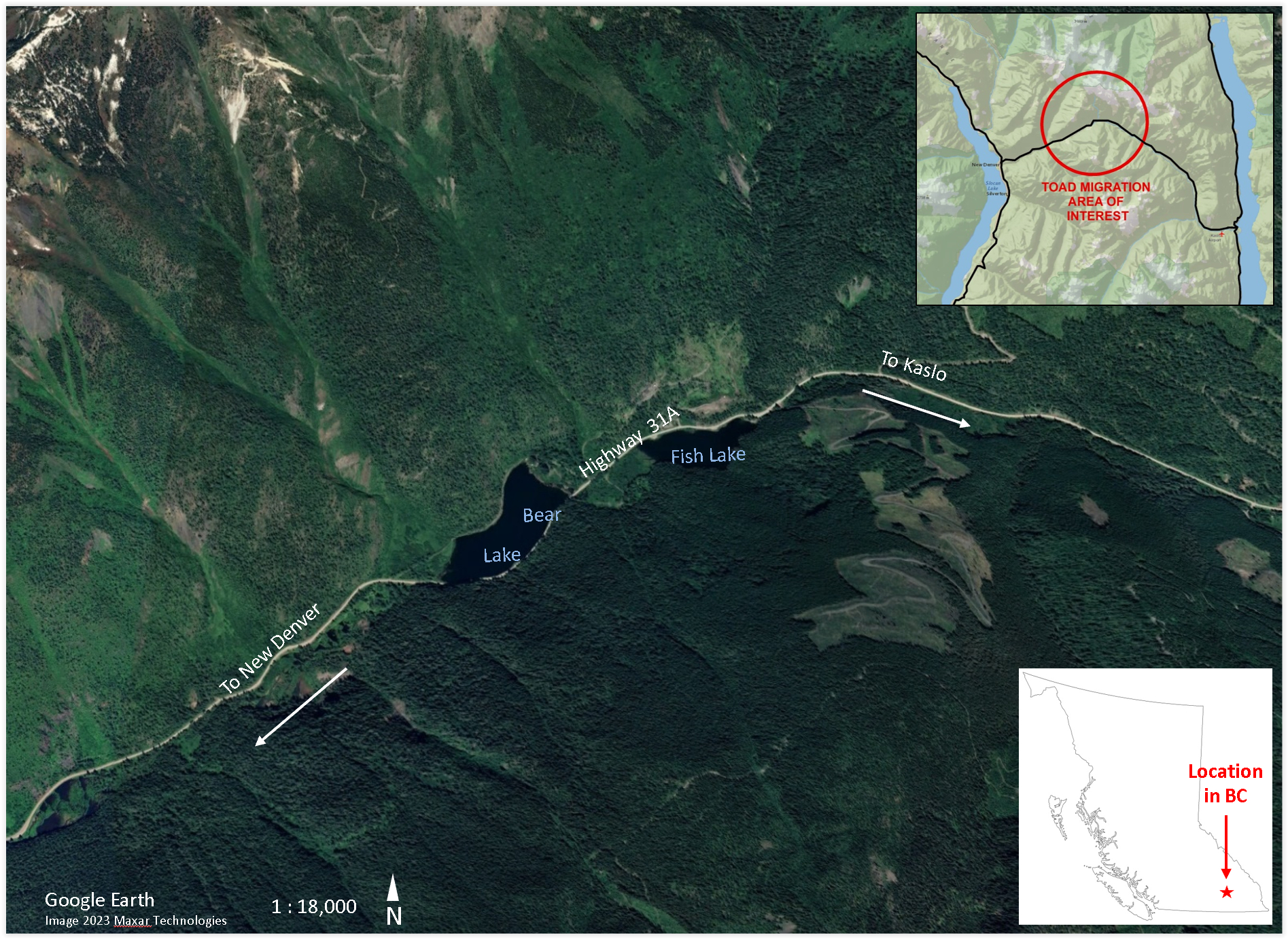}
    \caption{Fish and Bear Lakes study area is located in a
mountain pass within the BC Highway 31A corridor of the Central Selkirk Mountains. Fish
Lake provides the primary breeding and rearing habitat and
consequently the highest toad mortality occurs on the highway segment
directly adjacent to this lake. Adult toads migrating down from the
mountains on the north side of the highway are most at risk. 
Very few toads approach the lake from the south.  The area of interest (inset map: red circle) is approximately 7 km in diameter, and illustrates the potential maximum distance toads could travel between terrestrial foraging and hibernation areas and aquatic breeding sites~\citep{cosewic:2012}.}
    \label{fig:StudySiteSatellite}
\end{figure}

The study area is located at Fish and Bear Lakes along BC Highway 31A
in the Central Selkirk Mountains in southwestern British Columbia near
the townsite of Retallack and between the villages of New Denver and
Kaslo (Figure~\ref{fig:StudySiteSatellite}). The area encompasses a mountain pass at 1080~m comprised of two connected lakes; Bear Lake lies to the west and flows into Fish Lake, creating rich wetland habitat between the lakes.  Fish Lake provides exceptional toad breeding and rearing habitat along its
shallow north and west shores, and acts as the nursery for this
regional toad population. Although adult toads occasionally use
Bear Lake, we found no evidence of breeding there, likely due to
the steep-sided lakeshore, lake depth, cooler water temperatures, and
log jams along the shoreline. 

The terrestrial habitat to the north of the lake is of particular interest, as this habitat is favoured by the toads and separated from the lake by the highway.
Consequently, adult toads living north of Fish Lake must
cross the road twice during the annual breeding migration: Once to
reach the breeding area and again to return to terrestrial
habitat post-breeding. During late summer and fall, a third
migration occurs as new metamorphs (toadlets) emerge from Fish Lake
and attempt to cross the highway to reach terrestrial habitat where
they will spend the majority of their juvenile and adult lives (Figure~\ref{fig:life-cycle-diagrams}(a).

\subsection{Life Cycle and Migration Behaviour}
\label{sec:life+migration}

\begin{sidewaysfigure}
\centering
\includegraphics[width=\textwidth]{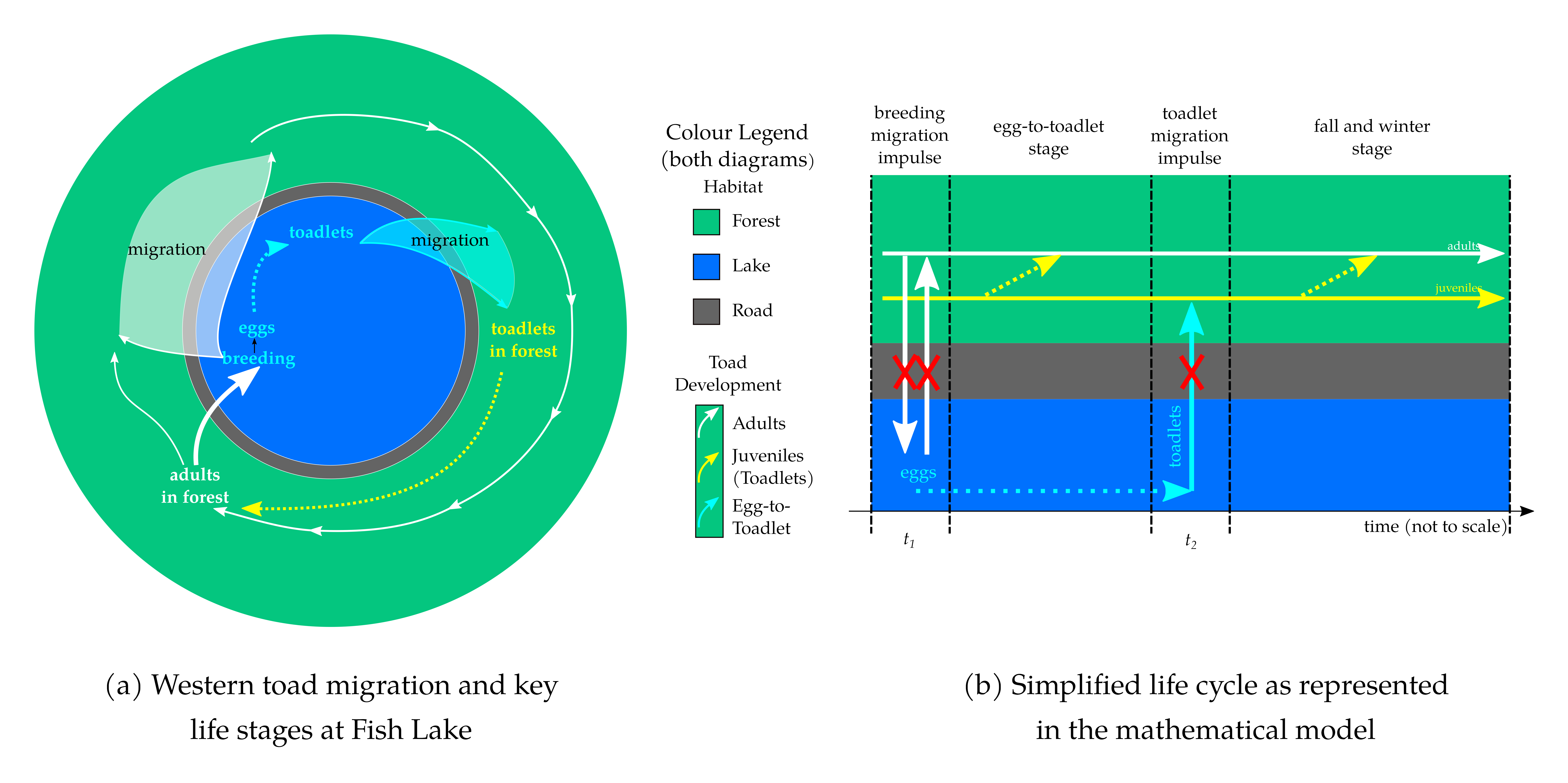}
\caption{
{\RRR Life cycle diagrams.  
{\bf (a)} Diagram showing the western toad migration and key life stages at Fish Lake.  In early spring (bottom left), adults migrate (thick white arrow), or not (thin white arrow) to the lake to breed.  Post-breeding, adults migrate back to the forest over several weeks (white shading, top left).  Adults remain in the forest until the next breeding season (outer white arrow).  Eggs in the lake mature (dotted cyan arrow) into toadlets who migrate to the forest in the last month of summer (cyan shading, top right).  In the forest, toadlets mature into adults (dotted yellow arrow).  Maturation takes several years (not shown).
{\bf (b)} Model simplification: Multi-week migrations are collapsed into single time points, i.e., impulse events (vertical arrows) modeled by equations~\eqref{eq:breeding-crossing} and~\eqref{eq:toadlet-crossing}.  Breeding adults migrate at $t_1$ and toadlets at $t_2$.  Red `X': points of potential road mortality.  Gradual processes (non-horizontal arrows) include maturation (dashed yellow arrows)) and mortality (white and yellow solid arrows).  These processes are represented by equations~\eqref{eq:toadlet} and~\eqref{eq:winter}.}  
}
\label{fig:life-cycle-diagrams}
\end{sidewaysfigure}

Mature adult toads live entirely in
terrestrial habitat except during a brief period in the spring when they migrate to aquatic habitat to breed. At our high elevation site, adults
migrate to Fish Lake from late April to early June, shortly following ice melt. Peak migration 
typically occurs from mid to late May.  Adult breeding toads must cross the highway to breed and lay eggs in the lake and, post-breeding, cross the highway again to return to the forest where they remain to forage and hibernate.  The post-breeding migration is heaviest immediately post-breeding, but carries on throughout the summer.  The eggs in the lake
develop into tadpoles that metamorphose into toadlets.  In late
summer, the toadlets leave the lake, crossing the road
to enter the terrestrial habitat, where they join the juvenile population. Juveniles remain in the forest as a non-breeding population for 2-6 years until they become sexually mature and return to the lake as adults to breed~\citep{cosewic:2012}. The cycle then repeats. 

{\RRR Figure~\ref{fig:life-cycle-diagrams}(a) illustrates the biphasic life cycle of western toads in our study area (note that Figure~\ref{fig:life-cycle-diagrams}(b) illustrates the simplified life cycle represented in the model - it is placed within Figure~\ref{fig:life-cycle-diagrams} for ease of comparison between the two diagrams, but is discussed in Section~\ref{sec:model}).  
 We represent the life cycle as a circle, to illustrate the fact that this life cycle repeats year after year. The road is also represented as a circle, as it remains a migration barrier, and source of road mortality, throughout the western toad life cycle.}

Extrapolating from other studies, we assume that female
toads mate 1-3 times over their lifespan \citep{BreedingFrequencyOregon},
with the majority of females only breeding once \citep{cosewic:2012}.  Breeding pairs will produce between 3,000 and 17,000 eggs \citep{maxell:2002, summitLake}. As is typical in amphibians however, only a small percentage of  eggs survive to the toadlet stage. The entire process of metamorphosis (from egg to tadpole to toadlet) at Fish Lake proceeds at a rate that is temperature-dependent but is generally completed within 3 months, with tadpoles hatching in 7-14 days. 

Metamorphosis from tadpole to toadlet at Fish Lake starts
generally near the end of July and continues to early September, with the peak
occurring in August. At the toadlet stage, the metamorphs have obtained
the final body shape of adult toads, and migrate across the highway to enter their terrestrial habitat.  These toads remain part of the juvenile population until they become reproductively active.  Sexual maturation occurs within 2-3 years for male toads, and 4-6 years for females \citep{cosewic:2012}.

The migration events of toads crossing Highway 31A follow diurnal patterns that differ for adults and toadlets.  Breeding adults most actively cross the highway in the evening, with peak activity levels between dusk and midnight.  
This period tends to coincide with less traffic volume than is typical
during the day. Gravid females, however, appear to linger on or near the warm asphalt, increasing their risk of being killed by vehicles, along with their eggs. 
Toadlet migrations take two forms, either mass migrations, generally associated with rain events, or gradual “trickle" migrations between rain events.  
These migrations occur during the day and in late summer, when traffic volumes are considerably higher.  Casualties can thus be significant, especially during mass migrations.

\section{Methods}
\label{sec:methods}


\subsection{Field Study}
\label{sec:data}

Data on adult toad presence, on and near the highway, {\RR vehicle presence,} and vehicle-induced mortality was collected annually from 2016-2021 at Fish and Bear Lakes. On spring and summer nights when breeding adult toads are most active, researchers and volunteers surveyed the area{\RR, recording western toad and vehicle numbers} on Highway 31A.  Surveys were also performed along adjacent recreation trails and nearby logging roads, but few\footnote{Too few toads were found to warrant further survey effort.} toads were found in these areas.

Researchers patrolled Highway 31A from dusk to midnight, recording sex, animacy (whether the animal is alive or dead), GPS location, orientation and direction of travel, and whether females were gravid (egg-bearing). {\RR All vehicles were counted, but not distinguished by type.}  
In addition to documenting timing, locations, and patterns of migrations, an important objective of the field surveys was to remove adult toads and toadlets from the highway and out of harm's way.

The data was gathered over several months from late spring (i.e., end of April or early May, depending on the timing of ice melt) to early fall (i.e., end of September) and is shown in Table~\ref{tbl:data}.  The highway experienced typical traffic levels in 2016--2019, and drastically reduced levels in 2020--2021 due to COVID-19 pandemic travel restrictions.

{\RR All work was done under the BC Ministry of Forests, Lands, Natural Resource Operations, and Rural Development guidelines and handling permit \#CB21-624388 and approved ethics guidelines.}

\subsection{Mathematical Model}
\label{sec:model}

Our model is based on our mechanistic understanding of the western toad system at Fish Lake, and the available data.  As with any modelling exercise, there is a trade-off between model complexity and the degree to which model behaviour can be understood and thus lead to predictive insights \citep{haefner:2005}. 
For example, from the point of view of population persistence, we only need to know how many toadlets make it across the road safely over the entire toadlet migration, and not the detailed specifics of the number of toadlets killed each day. 
We therefore simplify some of the life cycle details (chiefly, collapsing migration intervals to single time points) thereby focussing on the components that are most important to our study.
We also need to distinguish between two groups of metamorphs: Those that are in the lake and haven't yet crossed the highway, and those that have safely made it to terrestrial habitat.  We name the first group ``toadlets" and the second group ``juveniles" (while both groups are considered juveniles, this terminology is used here to distinguish pre- and post-crossing metamorphs).  A diagram illustrating the simplified life cycle represented in our model is shown in Figure~\ref{fig:life-cycle-diagrams}(b).  Below, we describe the model in detail.

\subsubsection{Model Definition}
\label{sec:modeldef}

We consider only the females of each stage since they determine the reproductive capacity of the population \citep{BreedingFrequencyOregon, hebblewhite:2003} and female road mortality is a critical factor in population persistence \citep{winton:2020}.
{\RR This approach is well established in population modelling, especially when density-dependent effects are minimal and can be reasonably excluded \citep{hostetler:2013, mollet:2002, halley:1996, crowder:1994, barbosa:2020}. Focussing on females allows for a simplified yet biologically meaningful framework for assessing population persistence.}
\begin{table}
\centering
\caption{Model variables (top) and parameters (bottom). The default values are either the midpoint of the range, the mean of the available data, or the most accepted value. For details, see Supplement~\ref{sec:paramVals}. $\zeta_g>1$ and $\zeta_x>1$ indicate that gravid females and toadlets are more likely to suffer highway-crossing mortality than non-gravid adult females. $\dag$``adults" refers to non-gravid adult female toads." $\ddag$ Note that we require $0<(1-\zeta_im_z)<1$ where $i\in\{g,x\}$, placing an additional restriction on the values of $\zeta_i$ and/or $m_z$. {\RR The parameters that are {\it rates} are the juvenile maturation rate $\delta$, and the three death rates $\mu_i$.}}
\begin{tabular}{lll}
\multicolumn{3}{l}{Variables} \\
\toprule
Symbol & Name   &  Life cycle interval \\ \midrule
$t$    & time (days) &  \\
$x(t)$ & toadlets (females only) & egg-to-toadlet (i.e., to 1st migration)\\
$y(t)$ & juveniles (females only) & 1st migration to sexual maturity \\
$z(t)$ & breeding adults (females only) & sexual maturity onward \\
\medskip \\
\multicolumn{3}{l}{Parameters} \\
\midrule
Symbol & Definition   &  Default Value \hfill (range) \\ \midrule
$m_z$  & highway crossing mortality of  & 10\% \hfill (8\%-50\%) \\
       & \quad adults$\dag$, per crossing, one way  & \\ 
$\zeta_g$ & highway crossing mortality ratio & 2$\ddag$ \hfill (1-3) \\
         & \quad gravid females : breeding adults & \\
$\zeta_x$ & highway crossing mortality ratio & 1.3$\ddag$ \hfill (1-2) \\
         & \quad toadlets : breeding adults & \\
$r$        & clutch size - female eggs & 6000 \hfill (3000-7000) \\
$\delta$ & juvenile maturation rate  & 0.0017 \hfill (0.0014-0.0021) \\
$\alpha$  & probability adult breeds more & 0.11 \hfill (0.053-0.24) \\
          & \quad than once & \\
$\mu_i$  & death rate of population $i$ & $\mu_x = 0.027$ \hfill (0.021-0.033) \\
         &                              & $\mu_y = 0.01$ \hfill (0.0041-0.021) \\
         &                              & $\mu_z = 0.0029$ \hfill (0.0016-0.0041) \\ 
$K$ & lake carrying capacity for & 2,000,000 \hfill (1,000,000 - 4,000,000) \\
 & \quad toadlets \\
 $T=t_2-t_1$ & days between spring and fall migrations & 90 \\
 $L$ & number of days in one year & 365 \\
\bottomrule
\end{tabular}
\label{tab:SymbolsTable}
\end{table}

The spring adult breeding and late summer toadlet migrations each occur over an extended period of time, however, peak movement occurs over an interval of 2-3 weeks.  This interval is short compared to the full year, so {\RR we approximate these migration periods as impulse events, which are sudden, discrete changes in an otherwise continuous population model (compare the two diagrams in Figures~\ref{fig:life-cycle-diagrams}.  Following similar work in \citet{BirthPulse1, BirthPulse2, BirthPulse3, BirthPulse4}, we use a stage-structured impulsive ODE model to track population dynamics across the three key life stages (egg-to-toadlets ($x(t)$), juveniles ($y(t)$), adults ($z(t)$)) with the impulsive component capturing instantaneous events like migration.  The three life stages correspond to the cyan, yellow, and white arrows in Figure~\ref{fig:life-cycle-diagrams}(b).}

The model has a period of 1 year or $L=$ 365 days with year number indicated by the index $n$.  
Each year has four behavioural periods: (1) The breeding impulse (at $t=t_1+nL$ days), (2) the egg-to-toadlet maturation interval (for $t\in(t_1+nL,t_2+nL)$ days), (3) the toadlet migration impulse (at $t=t_2+nL$ days), and (4) the fall and winter maturation period (for $t\in(t_2+nL,t_1+(n+1)L)$ days).  {\RR Mathematically, this structure translates to a model~\eqref{eq:model} with twelve equations: one for each life stage in each behavioural period.  Each equation is either an impulse equation or an ordinary differential equation (ODE).  Impulse equations describe the migrations, and ODEs describe the growth stages between the migrations.  For both equation types, the left hand side (e.g., $\Delta x$ or $\dot{x}$) represents the change in the population size.  For the impulse equations, this change is a sudden jump in the number of individuals as a result of migration.  For the ODEs, the change is gradual. The meaning of the terms on the right hand side is indicated in overbraces included with the equations.  Note that all individuals in the model are female, so, e.g., ``total eggs" means ``total female eggs", and ``adults" means ``female adults".}
Mathematically, we have
\allowdisplaybreaks
\beqsub
\label{eq:model}
\begin{align}
\phantom{{\text{Breeding Migration }}}
\mathllap{\text{\bf Breeding Migration }} & {\text{\bf Impulse: }} t=t_1+nL \nonumber \\
& \begin{aligned}
\mathllap{\Delta x(t)} & = \overbrace{B\left((1-\zeta_g m_z)z(t^-)\right)}^{\text{total eggs laid}} \overbrace{\left((1-\zeta_gm_z)z(t^-)\right)}^{\mathclap{\substack{\text{adults surviving} \\ \text{road crossing}}}} \\
\mathllap{\Delta y(t)} & = \overbrace{0}^{\text{no change}} \\
\mathllap{\Delta z(t)} & = \overbrace{\alpha}^{\mathclap{\substack{\text{fraction} \\ \text{to repeat} \\ \text{breeding}}}} \, \, \, \overbrace{(1-\zeta_gm_z)(1-m_z)\, z(t^-)}^{\mathclap{\substack{\text{adults surviving} \\ \text{both road crossings}}}} \\
\end{aligned}
\label{eq:breeding-crossing} \\
\nonumber \\
\mathllap{\text{\bf Egg to Toadlet Stag}} & {\text{\bf e: }} t\in(t_1 + nL, t_2 + nL) \nonumber \\
& \begin{aligned}
\mathllap{\dot{x}(t)} & = \overbrace{- \, \mu_x x(t)}^{\text{death}} \\
\mathllap{\dot{y}(t)} & = \overbrace{- \, \delta y(t)}^{\mathclap{\substack{\text{maturation} \\ \text{to adult}}}} \, \, \overbrace{- \mu_y y(t)}^{\text{death}} \\
\mathllap{\dot{z}(t)} & = \overbrace{\delta y(t)}^{\mathclap{\substack{\text{maturation} \\ \text{from} \\ \text{juvenile}}}} \, \, \overbrace{- \, \mu_z z(t)}^{\text{death}} 
\end{aligned}
\label{eq:toadlet} \\
\nonumber \\
\mathllap{\text{\bf Toadlet Migration I}} & {\text{\bf mpulse: }} t = t_2 + nL \nonumber \\
& \begin{aligned}
\mathllap{\Delta x(t)} & = \overbrace{- \, x(t^-)}^{\mathclap{\substack{\text{all} \\ \text{toadlets} \\ \text{migrate}}}} \\
\mathllap{\Delta y(t)} & = \overbrace{(1-\zeta_xm_z) x(t^-)}^{\mathclap{\substack{\text{toadlets} \\ \text{surviving} \\ \text{road crossing}}}} \\
\mathllap{\Delta z(t)} & = \overbrace{0}^{\text{no change}} 
\end{aligned}
\label{eq:toadlet-crossing} \\
\nonumber \\
\mathllap{\text{\bf Fall and Winter Sta}} & {\text{\bf ge: }} t\in(t_2 + nL, t_1 + (n+1)L) \nonumber \\
& \begin{aligned}
\mathllap{\dot{x}(t)} & = \overbrace{0}^{\text{no change}} \\
\mathllap{\dot{y}(t)} & = \overbrace{- \, \delta y(t)}^{\mathclap{\substack{\text{maturation} \\ \text{to adult}}}} \, \, \overbrace{- \, \mu_y y(t)}^{\text{death}} \\
\mathllap{\dot{z}(t)} & = \overbrace{\delta y(t)}^{\mathclap{\substack{\text{maturation} \\ \text{from} \\ \text{juvenile}}}} \, \, \overbrace{- \, \mu_z z(t)}^{\text{death}} \\
\end{aligned}
\label{eq:winter}
\end{align}
\eeqsub
Note that $t^-$ denotes the moment immediately before time $t$. As in \citet{BirthPulse1}, the birth function $B(\cdot)$, or per female production of female offspring, is a real-valued function that maps the number of female adults to the number of female eggs produced each year ($B: \mathbb{R^+} \to \mathbb{R^+}$).

The model variables and parameters are defined in Table~\ref{tab:SymbolsTable}.  Details are given in Appendix~\ref{sec:paramVals}.  The spring breeding and toadlet migration impulses occur at times $t_1$ and $t_2$, respectively. Highway-crossing mortality for non-gravid adult female toads is $m_z$. Gravid females and toadlets may have increased mortality relative to non-gravid adult females (see Section~\ref{sec:life+migration}), and so gravid female and toadlet mortalities are $\zeta_gm_z$ and $\zeta_xm_z$, respectively, where $\zeta_g>1$ and $\zeta_x>1$.

\subsubsection{Mathematical Analysis}
\label{sec:analysis}

Since model~\eqref{eq:model} is linear, we can obtain an exact analytical solution for the population vector $(x(t),y(t),z(t))$ (see Supplement~\ref{sec:modelSol} for the solution and its derivation).

We are interested in the periodic steady state solutions of~\eqref{eq:model}, obtained as $n\rightarrow\infty$ (see Supplement~\ref{sec:NEWPerSteadyStatesDerivation} for the derivation).  We choose to focus on the population vector at the time of the breeding and toadlet migrations.  We define the population sequences 
\begin{align*}
x_1^n & = x(t_1 + nL), n \in \mathbb{N} \cup \{0\}, \\
x_2^n & = x(t_2 + nL), n \in \mathbb{N} \cup \{0\},
\end{align*}
where the population is evaluated at the {\it{end}} of each impulse.  For simplicity, we refer to the egg-to-toadlet population $x(t)$ as the ``toadlet" population.  Similarly, we define $y_{t_1}^n$, $y_{t_2}^n$, $z_{t_1}^n$, and $z_{t_2}^n$ for the juvenile and adult populations. The limits of these sequences, when they exist, are the long-term periodic steady states of the system at the migration times.  We denote these limits by $(x_i^*, y_i^*, z_i^*)$ where $i\in\{1,2\}$.   

%
Solving for the steady state solutions (see Supplement~\ref{sec:NEWPerSteadyStatesDerivation} for details), we obtain
\beq
B\left( \frac{z_1^*}{\alpha(1-m_z)} \right) = \frac{\alpha(\delta+\mu_y-\mu_z)}{Q(1-\zeta_xm_z)\delta} 
\left[ 
\frac{1}{\alpha(1-\zeta_gm_z)} - (1-m_z)e^{-\mu_zL}
\right]
\label{eq:NEWz1star_roots}
\eeq
where
\beq
Q = \frac{e^{-\mu_xT}}{1-e^{-(\delta+\mu_y)L}}
\left[
e^{-\mu_z(L-T)}\left( 1 - e^{-(\delta+\mu_y)L} \right)
+ e^{-(\delta+\mu_y)(L-T)}\left( e^{-\mu_zL}-1 \right)
\right]
\label{eq:Q}
\eeq
with $T=t_2-t_1$.  
When the right hand side of~\eqref{eq:NEWz1star_roots} is positive and less than the maximum value of the birth function $B(\cdot)$, there is at least one positive solution $z_{t_1}^*$. In this paper, we take $B(z)$ to be the Beverton-Holt growth function \citep{BirthPulse2, kot:2001}, adjusted for the fact that $z$ is just the female half of the population (see Appendix~\ref{sec:BHgrowth}).  We arrive at
\beq
\label{eq:BevertonHoltBirth}
B(z) = \frac{rK}{K + (r-1)\,2z}, \qquad r \geq 0,
\eeq
where $K$
is the carrying capacity for toadlets (male and female, i.e., the entire population) at Fish and Bear Lakes, $r$ is the average number of female eggs per female toad, and we assume that there is a 1:1 ratio of male to female offspring and so the increase in the female population is half the total increase.  

With $B(z)$ specified, we can determine the unique positive $z_{t_1}^*$ explicitly from~\eqref{eq:NEWz1star_roots}.  We obtain
\beq
z_1^* = \frac{K\alpha(1-m_z)}{2(r-1)} 
\left(
\frac{Qr\delta}{\delta+\mu_y-\mu_z} 
\frac{(1-\zeta_x m_z)(1-\zeta_gm_z)}{1-\alpha(1-\zeta_gm_z)m_ze^{-\mu_zL}}
-1
\right)
\label{eq:NEWz1star}
\eeq
{\RRR This equation is the analytic solution of our model.}
Note that, for~\eqref{eq:NEWz1star} to make sense, we require that
\beq
0<(1-\zeta_g m_z)<1 \quad {\text{ and }} \quad 0<(1-\zeta_x m_z)<1,
\label{eq:zeta-constraint}
\eeq
which places a constraint on the allowable values of $\zeta_g$ and $\zeta_x$ for each value of $m_z$.  We can address this constraint in a straightforward manner by expressing $z_{t_1}^*$ in terms of adult female survivorship, $s_z=1-m_z$, and factor reducing survivorship $\eta_g$ and $\eta_x$ where 
\beq
\eta_g s_z=1-\zeta_g m_z \quad {\text{ and }} \quad \eta_x s_z=1-\zeta_x m_z.
\label{eq:survivorship-translation}
\eeq
Since $0<s_z<1$ and $0<\eta_i<1$, $i\in{g,x}$, we can be certain that the factors on the left in the constraints~\eqref{eq:survivorship-translation} will always be nonnegative.  See the supplementary material, Section~\ref{sec:NEWPerSteadyStatesDerivation}, for the survivorship version of~\eqref{eq:NEWz1star}.

%
To assess how changes in highway-crossing mortality affect population persistence, we focus on that point in the annual life cycle where the toad population is at its minimum.  The minimum in the adult population occurs immediately after the spring migration.  We thus have
\beq
\label{eq:zstar_zmin}
z_{min} = z_1^*.
\eeq
We are also interested in the total number of potential breeding toads, that is, the number of adult female toads immediately preceding the spring migration.  This value is obtained from $z_{t_1}^*$ by reversing the two spring highway crossings, and is given by 
\beq
\label{eq:NEWzstar_max}
z_{max} = \frac{z_1^*}{\alpha(1-\zeta_gm_z)(1-m_z)}
\eeq
%
%
where we use the subscript $max$ to indicate that this value is the maximum possible number of adult female breeding toads, i.e., the number that would breed were highway-crossing mortality equal to 0. 

Finally, we compute the mortality level at which extirpation occurs by solving for $m_z$ in~\eqref{eq:NEWz1star} when $z_1^*=0$, which yields
\beq
\label{eq:NEWmzc}
m_z^c = \frac{b\pm\sqrt{b^2-4ac}}{2a}
\eeq
%
%
where the coefficients $a$, $b$, and $c$ are given by
\beqsub
\beqa
a & = & \zeta_g \left( Qr\delta\zeta_x - \alpha(\delta+\mu_y-\mu_z)e^{-\mu_zL} \right) \\ 
b & = & (\zeta_x+\zeta_g) - \alpha (\delta+\mu_y-\mu_z) e^{-\mu_zL} \\
c & = & Qr\delta - (\delta+\mu_y-\mu_z)
\eeqa
\label{eq:mc-coeffs}
\eeqsub
the critical mortality level is the smallest positive solution from~\eqref{eq:NEWmzc}.  Note that the critical mortality level in~\eqref{eq:NEWmzc} does not depend on the carrying capacity $K$.  
See Appendix~\ref{sec:extirpation} for the critical mortality level in terms of survivorship.
We will use the quantities~\eqref{eq:zstar_zmin}, \eqref{eq:NEWzstar_max}, and~\eqref{eq:NEWmzc} as indicators of toad population success.

\subsubsection{Sampling Across Parameter Space}
\label{sec:sampling}

We addressed the uncertainty around parameter values (see the ranges listed in Table~\ref{tab:SymbolsTable}) using a Monte-Carlo approach.  We assumed each parameter to be uniformly distributed across the range given in Table~\ref{tab:SymbolsTable}, and randomly selected $10^5$ points within the full parameter space.

\subsubsection{Endangerment Thresholds}
\label{sec:endanger-def}

{\RRR For our steady state population predictions, we are interested in determining the status of the population.  According to COSEWIC, a population that is ``Threatened" comprises 1000 individuals, and one that is ``Endangered" comprises 250 individuals \citep{COSEWICendangered}.  As our model is restricted to the female half of the population, the predicted steady state population is considered threatened if $z_{min}<500$, and endangered if $z_{min}<125$.  }

\section{Results}
\label{sec:results}

\subsection{Field Study}
\label{sec:field-study-results}

In each year of the study except the first, the percentage of female mortality is higher than that of male mortality (Table~\ref{tbl:data}).  In the non-COVID years, female mortality ranges from 10\% to 25\%, with gravid females representing 45\%-54\% of the dead female toads.  This last result suggests that gravid and non-gravid females are equally likely to suffer from road mortality, but calculations determining monthly mortality show that gravid females do indeed suffer higher mortality than non-gravid females, particularly in May (see the calculations for $\zeta_g$ in Supplement~\ref{sec:paramVals}).   
In addition, the gravid females were observed spending more time on the road than non-gravid females.  We hypothesize that these females are seeking the warmth from the asphalt, as this helps them digest the larger quantities of food they need to consume for egg development.  Asphalt warmth may be a key resource for this high elevation population.  This behaviour may explain the higher mortality of females as compared to males.   
\begin{table}
\centering
\caption{Adult toad highway-crossing data gathered at the Fish and Bear Lakes study area between the months April-September.  In the ``females", ``males", and ``total" columns, the data given is number of dead toads : number of live toads, and (dead toads as a percentage of dead$+$live toads). In the ``gravid" column, the data given is the number of dead gravid females : number of dead females, and (dead gravid females as a percentage of dead female toads). {\RR The ``total" column includes toads for which sex was not identified.}  $^*$Mortality in 2020 and 2021 is significantly lower due to the COVID-19 pandemic travel restrictions.  
} 
\begin{tabular}{l|rrr|r}
Year & females & males & total & gravid \\ \hline
2016 & 24:73 (25\%) & 26:50 (34\%) & 93:127 (42\%) & 14:24 (58\%) \\
2017 & 29:251 (10\%) & 18:222 (8\%) & 72:484 (13\%) & 13:29 (45\%) \\
2018 & 38:252 (13\%) & 23:301 (7\%) & 81:561 (13\%) & 17:38 (45\%) \\
2019 & 33:161 (17\%) & 42:283 (13\%) & 89:454 (16\%) & 15:33 (45\%) \\
\hline
2020$^*$ & 6:117 (5\%) & 1:138 (1\%) & 7:265 (3\%) & 2:6 (33\%) \\
2021$^*$ & 6:98 (6\%) & 5:93 (5\%) & 14:191 (7\%) & 3:6 (50\%) \\
\hline
\end{tabular}
\label{tbl:data}
\end{table}

We fit several generalized linear regression models, assuming a Poisson distribution, for (1) the number of vehicles on the highway, (2) the number of live toads on the highway, and (3) the number of dead toads on the highway during the collection period (see Supplement~\ref{sec:regression} for full analysis). Sexing dead non-gravid toads was often difficult, so the full toad population was used for this analysis. 
All models were fit with a random effect of the year and an offset for the number of minutes per survey period. We used the Akaike Information Criterion (AIC) to compare models.

\begin{figure}
    \centering  \includegraphics[width=0.9\textwidth]{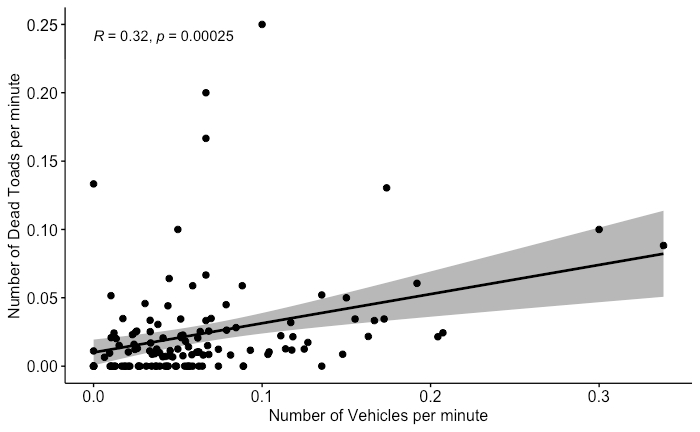}
    \caption{Scatter plot of adult toad mortality rate per minute (all toads, i.e., male and female (gravid and non-gravid)) as a function of the number of vehicles on the road per minute. The fitted line shows a linear regression between these two variables, as well as the Pearson correlation ($R$) and the significance level ($p<0.05$). This plot shows the simplest model possible illustrating the relationship between the number of vehicles per minute and the number of dead toads per minute. The full analysis (Supplement~\ref{sec:regression}) gives estimates for all of the model coefficients.}
    \label{fig:vehpermin-vs-mortality}
\end{figure}

The best fitting model for (1) the number of vehicles on the highway, has explanatory variables of month (category), start time (category - evening or night) and whether the year preceded or was during COVID (indicator - before 2020 or not).  The best fitting model for (2) the number of live toads present, has the explanatory variables of month, the number of vehicles present, and whether the year preceded or was during COVID. Finally, the best fitting model for (3) the number of dead toads present, has the explanatory variables of number of live toads present and number of vehicles on the highway, with both having significant impact ($p<0.05$). With this model we estimate that, for each additional vehicle on the highway, there is a $3.1\%\pm1.3\%$ ($p=0.020$) increase in the number of dead toads.  The model fit (3) and data are shown in Figure~\ref{fig:vehpermin-vs-mortality}.

\subsection{Mathematical Model Results}
\label{sec:model-results}

We now use the mathematical model to predict how highway mortality levels, both higher and lower than those currently observed, will affect the western toad population at Fish and Bear Lakes.

\subsubsection{Population vs. highway-crossing Mortality}
\label{sec:results-popn}

\begin{figure}
    \centering
    \includegraphics[width=0.8\textwidth]{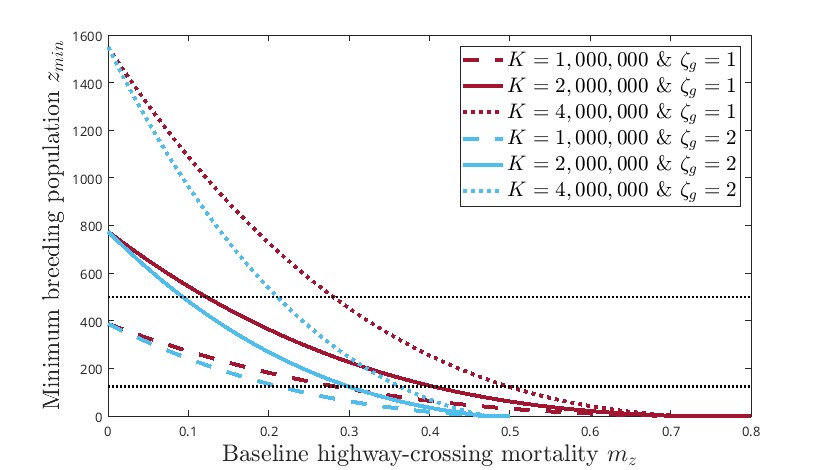}
    \caption{Plot of $z_{min}$ as a function of highway-crossing mortality (per crossing, ranging from 0 to 0.8 (80\%)) for breeding adults, and for different values of the carrying capacity $K$ and increased mortality factor for gravid females $\zeta_g$ (gravid vulnerability).  The values of $K$ are indicated by line type (dashed for $K=3,000,000$, solid for $K=2,000,000$, and dotted for $K=1,000,000$), and the values of $\zeta_g$ are indicated by colour (red for $\zeta_g=1$, blue for $\zeta_g=2$, colour online).    
    The upper and lower dotted black horizontal lines indicate, respectively, the ``threatened" and ``endangered" thresholds (see Section~\ref{sec:endangerment}).  
    The extirpation threshold $m_z^c\approx0.7$ for $\zeta_g=1$, and $\approx0.5$ for $\zeta_g=2$.  
    Parameter values are at the default values listed in Table~\ref{tab:SymbolsTable}. 
}
    \label{fig:maxminBreedingPop}
\end{figure}

We find that the lake carrying capacity and the gravid female mortality factor both have a strong effect on the predicted minimum breeding population level (Figure~\ref{fig:maxminBreedingPop}).  The former is very difficult to measure, and so we consider a wide range of values.
To interpret this plot, consider the following example: If the baseline highway-crossing mortality for female toads is 30\%, and the lake carrying capacity for females is $K=2,000,000$, the solid red and blue curves indicate that the minimum breeding population is approximately 250 females, if there is no additional mortality factor for gravid females (solid red curve), or 200 females, if gravid females are killed by vehicles at twice the rate of non-gravid females (solid blue curve).  At the mortality levels observed in 2016-2019, i.e, $m_z=0.1$ to $0.2$, the minimum breeding population only remains above the ``threatened" threshold for the higher two values of the carrying capacity (solid and dash-dot curves).

As the baseline highway crossing mortality $m_z$ increases, $z_{min}$  decreases rapidly.  Mathematical extirpation occurs when $z_{min}=0$. From Figure~\ref{fig:maxminBreedingPop}, we see that the extirpation point does not vary with $K$, and is approximately $m_z^c=0.7$ for $\zeta_g=1$ and $m_z^c=0.5$ for $\zeta_g=2$.  From Equation~\eqref{eq:NEWz1star}, we can find extirpation for highway-crossing mortality levels as low as 20\%, and the likelihood of extirpation increases quickly for higher levels of highway-crossing mortality (see Appendix~\ref{sec:extirpation}).  Real extirpation, however, will occur sooner, i.e., when the minimum breeding population is below some lower threshold (see Section~\ref{sec:endangerment}) and vulnerable to random fluctuations.  

\subsubsection{Endangerment}    
\label{sec:endangerment}

Using the Monte-Carlo sampling process (Section~\ref{sec:sampling}), we determine the proportion of parameter sets leading to a steady state population with a status of threatened, endangered, or neither.

\begin{figure}
    \centering
    \begin{subfigure}{0.45\textwidth}
       \centering
       \includegraphics[width=\textwidth]{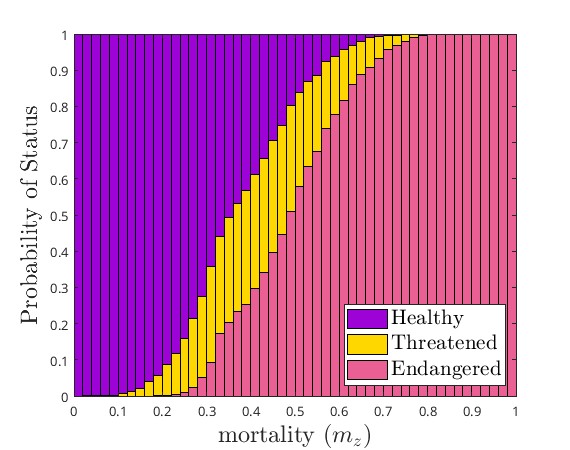}
       \caption{$K=1,000,000$}
       \label{fig:sustSnK1}
    \end{subfigure}
    \begin{subfigure}{0.45\textwidth}
       \centering
       \includegraphics[width=\textwidth]{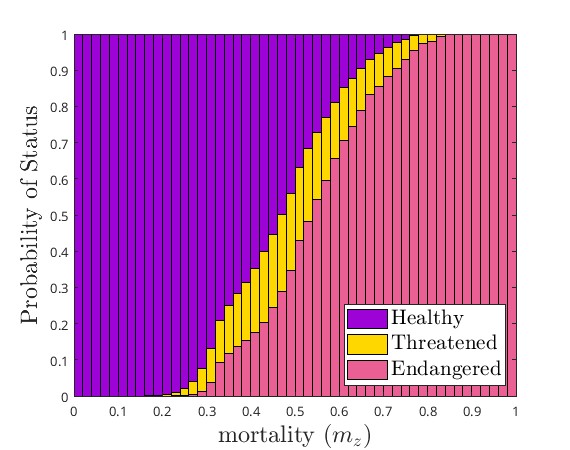}
       \caption{$K=3,000,000$}
       \label{fig:sustSnK2}
    \end{subfigure}
    \caption{Relative risk of the toad population reaching a status of endangered (red),  threatened (yellow), or healthy (purple) (colour online) as a function of baseline adult highway-crossing mortality ($m_z$), ranging from 0 to 1 (100\%).
    The height of the bar at each level of mortality corresponds to the proportion of parameter sets, at that level of mortality, resulting in a minimum steady state population falling within the appropriate interval ($z_{min}\leq125$ (red), $125<z_{min}\leq500$ (yellow), $z_{min}>500$ (purple), colour online). 
    Results are shown for (left) $K=1,000,000$ and (right) $K=3,000,000$.  For the remaining parameter values, the parameter space was sampled 100,000 times, using a Monte Carlo sampling of parameter values (Section~\ref{sec:sampling}) distributed uniformly over the ranges defined in Table~\ref{tab:SymbolsTable}.}
    \label{fig:sustSn}
\end{figure}

The results for $K=1,000,000$ and $K=3,000,000$ are shown in Figure~\ref{fig:sustSn}. At each value of $m_z$, the coloured histograms (colour online) show what proportion of parameter sets result in a population that is healthy (green), threatened (yellow), or endangered (red). To interpret these plots, consider the following example:  Let the baseline observed highway-crossing mortality value
be {\RRR $m_z=20\%=0.2$}.  Then, for $K=1,000,000$ (left sub-plot),
approximately 95\% of parameter sets result in a healthy population,  5\% result in a threatened population, and none of the parameter sets result in an endangered population.  At $m_z=0.4$, 40\% of parameter sets result in a healthy population, 20\% in a threatened population, and nearly 40\% in an endangered population.

We focus on the effect of the baseline mortality factor, $m_z$, as it can be directly controlled.  
Ideally, vehicle-caused mortality is nearly eliminated altogether by, e.g., installing toad tunnels under the highway.  In this scenario, we assume that $m_z<10\%$, and the proportions of parameter sets leading to healthy, threatened, and endangered populations changes to more than 95\%, less than 5\%, and 0\% for $K=1,000,000$.  At $K=3,000,000$ and less than 10\% highway-crossing mortality, all parameter sets lead to a healthy population.  From these two baselines, the relative likelihood\footnote{Note that ``relative likelihood" of some outcome (e.g., population persistence) means the "proportion of parameter sets" predicting that outcome.} of population persistence drops steeply as traffic and $m_z$ increase. In particular, the relative likelihood of population persistence at healthy levels drops below 50\% at a highway-crossing mortality of approximately 35\% or 40\% (i.e., at $m_z\approx0.35$ and $m_z\approx0.4$) for $K=1,000,000$ and $K=3,000,000$, respectively.  
The highest overall highway-crossing mortality observed in the data was over $40\%$.
For both values of the carrying capacity, as highway-crossing mortality increases there is a sharp transition from mortality levels mostly resulting in a healthy population, to mortality levels mostly resulting in an endangered population (Figure~\ref{fig:sustSn}).

This sharp transition is consistent with the mortality levels observed in the data.  The maximum traffic levels observed before and during the COVID pandemic were 0.12 and 0.09 vehicles per minute, respectively, while
maximum mortality levels before and during COVID were 23\% and 9\%, respectively.
Thus, while the non-COVID traffic levels were 25\% higher than during-COVID traffic levels, mortality was 150\% higher, meaning that a small increase in vehicles per minute can result in a large increase in mortality.

\begin{figure}
    \centering
    \begin{subfigure}{0.8\textwidth}
    \centering
      \includegraphics[width=\textwidth]{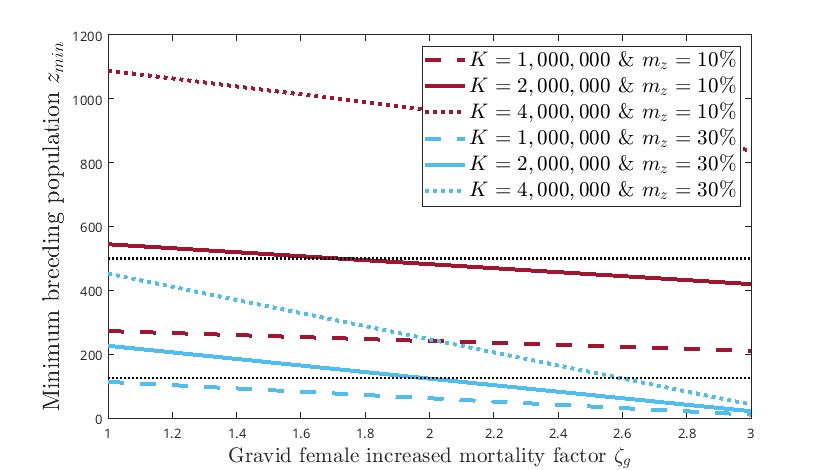}
      \caption{Minimum breeding population $z_{min}$ as a function of gravid female increased mortality factor $\zeta_g$. }
      \label{fig:sustSn-zeta-zmin}
    \end{subfigure}
    \begin{subfigure}{0.45\textwidth}
      \includegraphics[width=\textwidth]{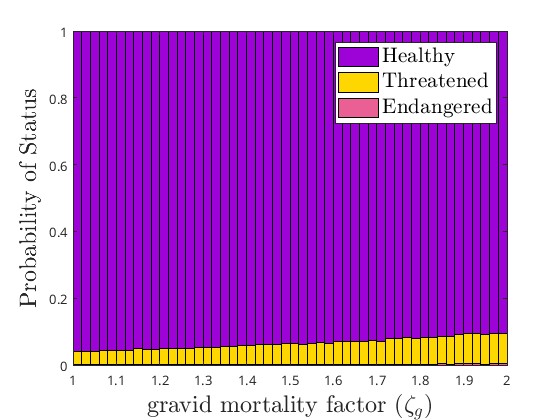}
      \caption{Risk of endangerment with maximum highway-crossing mortality $m_z=10$\%-$30$\%.}
      \label{fig:sustSn-zeta-threat1}
    \end{subfigure}
    \begin{subfigure}{0.45\textwidth}
      \includegraphics[width=\textwidth]{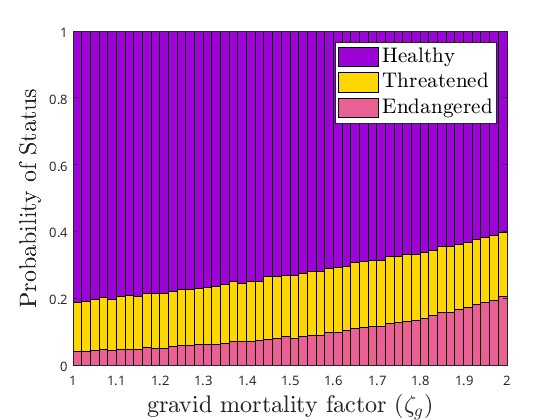}
      \caption{Risk of endangerment with highway-crossing mortality $m_z=10$\%-$50$\%.}
      \label{fig:sustSn-zeta-threat2}
    \end{subfigure}
    \caption{Plots showing the effect of the gravid female mortality factor $\zeta_g$ on population persistence.  (\subref{fig:sustSn-zeta-zmin}) Plot of $z_{min}$ versus $\zeta_g$, for different values of the carrying capacity $K$ (dot: $K=4,000,000$; solid: $K=2,000,000$; dashed: $K=1,000,000$) and non-gravid adult highway-crossing mortality $m_z$ (red: $m_z=10$\%; blue: $m_z=30$\%, colour online).  The dotted black horizontal lines  indicate the ``threatened" and ``endangered" thresholds (see Section~\ref{sec:endangerment}).  
      (\subref{fig:sustSn-zeta-threat1})--(\subref{fig:sustSn-zeta-threat2})  Risk of the toad population reaching a status of endangered (red),  threatened (yellow), or neither (purple) (colour online) as a function of $\zeta_g$.  The height of each coloured bar gives the proportion of parameter sets, at that value of $\zeta_g$, for which the population has the corresponding endangerment status. 
      Results are shown for $K=1,000,000$, and adult highway-crossing mortality $m_z$ of 10\%-30\% (left) and 10\%-50\% (right).  For the remaining parameter values, the parameter space was sampled $100,000$ times, using a Monte Carlo sampling of parameter values (Section~\ref{sec:endangerment}) distributed uniformly over the ranges defined in Table 1.}
    \label{fig:sustSn-zeta}
\end{figure}

The gravid female increased mortality factor is not directly accessible to management efforts, but is an unknown yet important factor in the predicted outcomes (Figure~\ref{fig:sustSn-zeta-zmin}).  
As expected, minimum breeding population size decreases with increasing $\zeta_g$, and with increasing $m_z$ (baseline adult highway-crossing mortality).  The observed dependence on $\zeta_g$ is linear.  For the parameter values shown, the minimum breeding population drops below the threshold for classification as ``threatened" for values of $\zeta_g$ as low as 1 (when $K=1,000,000$ and $m_z=0.1$, or for all three values of $K$ when $m_z=0.3$).  When highway-crossing mortality is only 10\%, the population does not drop to the level of ``endangered" for any gravid mortality factor shown.  When highway-crossing mortality rises to 30\%, and for all values of the carrying capacity, the population eventually enters the ``endangered" zone at some level of increased gravid mortality factor.  Indeed, for the lowest carrying capacity, the population is ``endangered" for all values of the gravid mortality factor shown, including $\zeta_g=1$. 

The probability of endangerment also increases with increasing $\zeta_g${\RR, though the increase is small (from 5--10\%) for mortality levels $m_z$ between 10\% and 30\%  (Figure~\ref{fig:sustSn-zeta-threat1}).  At higher mortality levels ($m_z$ between 10\% and 50\%), increased female mortality becomes a much more important factor (Figure~\ref{fig:sustSn-zeta-threat2}), with probability of endangerment ranging from 20--40\% as gravid mortality increases.  Our results suggest that the increased mortality of females may not be a problem at current traffic levels, but will become an important factor if there is a sufficient increase in traffic.}

{\RR Note that the histograms in Figure~\ref{fig:sustSn} and~\ref{fig:sustSn-zeta} are not directly comparable, as each one is plotting a different subset of the simulation results.  In particular, endangerment is less likely in the Figure~\ref{fig:sustSn-zeta} histograms because the data plotted here do not include $z_{min}$ values calculated for higher baseline mortality levels.  What we learn from Figure~\ref{fig:sustSn} is that small increases in baseline highway-crossing mortality level can result in a dramatic increase in the likelihood of endangerment if $m_z$ is close to the threshold value (where the yellow (colour online) component of the histogram rises rapidly).  From Figure~\ref{fig:sustSn-zeta} we learn that the increased vulnerability of gravid females has little effect on the likelihood of population endangerment until baseline mortality increases past 30\%.}


\section{Discussion}
\label{sec:discussion}


In this paper, we combine a western toad field study with a deterministic mathematical modelling study to understand the effect of highway-crossing mortality on population persistence.  {\RR This two-pronged approach is relatively uncommon, and is a powerful framework for wildlife management.} The data provide us with strong evidence that highway-crossing mortality increases substantially with increased traffic levels, and indicates appropriate values for the mortality parameters in our model.  Our mathematical model~\eqref{eq:model} provides
an estimate of both the qualitative and quantitative relationships between population size and highway-crossing mortality.  To obtain these relationships, we present a simplified version of the toad life cycle and the accompanying mechanistic mathematical model.  We emphasize here that our model is not meant to be a precise predictor of the population size for a given level of vehicle traffic, but rather a first step in understanding the level of extirpation risk posed by increased traffic. 

We are able to obtain explicit equations for the periodic steady state solutions of the model, and thus can calculate the steady state population during each part of the life cycle. In particular, we can calculate the extirpation threshold $m_z^c$ as a function of either adult highway-crossing mortality $m_z$ or gravid female increased highway-crossing mortality factor $\zeta_g$.  We observe a decrease in population size as highway-crossing mortality increases, and find that, for our default parameter values, the mathematical extirpation threshold is approximately $m_z^c=90$\% (Figure~\ref{fig:maxminBreedingPop}).  In practice, extirpation is likely to occur at highway-crossing mortality levels considerably lower than $m_z^c$.

Estimates of expected equilibrium values of the western toad population at Fish and Bear Lakes are shown for all possible levels of road mortality (from 0\% to 100\%) and across the range of plausible life history parameter values.  We find that the minimum female breeding population level is highly sensitive to increases in highway-crossing mortality, dropping rapidly as highway-crossing mortality increases. Furthermore, for smaller values of the carrying capacity $K$, the minimum female breeding population drops below the critical threshold for classification as ``threatened" \citep{COSEWICendangered} when highway-crossing mortality is still as low as 25\%.  Indeed, for $m_z=10$\%, the current estimate, the minimal female breeding population is estimated to be only 2-4 times the ``threatened" threshold (Figure~\ref{fig:maxminBreedingPop}).  We see that the carrying capacity $K$ is a key parameter, and an accurate estimate of $K$ is needed for precise predictions of population size.  {\RR This result suggests that it would be worthwhile to investigate other possible growth functions, to ensure that our results are robust across plausible growth functions.}

Due to the limited amount of population data, we obtained solutions using parameter values sampled across wide ranges of possible values.  These ranges were constructed from data or, when data were unavailable, through expert opinion and trends observed in the literature. We assumed{\RR, for simplicity,} that each parameter had a uniform distribution across the identified range.  {\RR Further field study data is needed to determine more precise ranges and distributions for the model parameters.} Nonetheless, when we repeated our calculations for normal parameter distributions centred at the middle of each range, we found changes in the critical values of highway-crossing mortality, but not in the general pattern of results.

Our sampling of the solution space is an initial attempt at determining the population's probability of persistence and vulnerability to increased traffic.  
Our work indicates that even if the population appears healthy, it may be closer to critical thresholds than current population counts would indicate. Our results are thus a call for further study of this and other western toad populations, especially in areas like the Highway 31A corridor where there is significant development and recreational pressures \citep{govBCimap:2021} that could put populations under threat.

The field research was well underway before the modelling study was conceived, thus the former was not specifically designed to provide data for the latter.  There are consequently two aspects of the field study that make estimation of model parameter values somewhat tricky.  First, when possible, toads were moved off the road and out of harm's way, which reduced highway-crossing mortality.  Second, our estimate of the total population is based on highway count data, though a small fraction of the population is able to avoid crossing the highway by moving east along the creek at the east end of Fish Lake and then directly south from the lake into surrounding terrestrial habitat. 
The field study thus gives a strong indication of the relationship between highway-crossing mortality and vehicle traffic levels, and the data provide reasonable first order estimates of model parameter values. {\RR In terms of model predictions, the uncertainty in parameter values
means that the tolerable degree of increased traffic may be higher than we predict.  Unchanged, however is our main result that the population exhibits a rapid switch from healthy to endangered as highway-crossing mortality increases past a certain threshold.}

Under current traffic volumes in the Highway 31A corridor, mortality per highway-crossing is estimated to be fairly low, ranging between 4\% and 20\% for adult toads.  Given that the observed highway-crossing mortality increased anywhere from 2 to 10 times between non-COVID and COVID years, we conclude that any development in the Highway 31A corridor will likely result in significant increases in highway-crossing mortality for western toads.  Highway mitigation infrastructure such as amphibian underpasses and fencing can be an effective means of preventing highway-crossing mortality and allowing continued migration of adult toads and toadlets \citep{BCMOE:2020, roadMortMitigation, jolivet:2008, soanes:2024}. 

While we have focussed on the effect of highway-crossing mortality on population persistence, anthropogenic activity in the neighbouring terrestrial habitat also negatively affects the western toad population at Fish and Bear Lakes.  These activities include clearcutting on crown and private land, destruction of riparian and wetland habitat, and increased human recreational use causing displacement and direct mortality of toads during all life stages.  All of these factors amplify the negative effects of highway-crossing mortality.  {\RR Future field and modelling work should consider these broader factors, as well as the additional threats imposed by climate change.}

The western toad population that we studied is one of a multitude of amphibian populations across the globe that are potentially at risk of extirpation due to road crossing mortality \citep{BCMOE:2020, roadMortMitigation, fahrig:1995, glista:2008, silva:2021, coelho:2012}.  These populations are an important component of healthy ecosystems, making their decline alarming, even more so in the context of our result that the transition from healthy to endangered can result from only a small increase in road crossing mortality. Field studies of road mortality and the effectiveness of mitigation measures offer possible solutions \citep{marcelino:2025, pinto:2024, hamer:2023, testud:2020} and reasons for hope \citep{moor:2022}.  Models such as the one presented here, that are also coupled with field data, can be an important tool to rigorously determine where to place mitigation measures, and which populations are most at risk \citep{lee:2022, petrovan:2019}. 
{\RRR One of the key advantages of our model is that we are able to provide an analytical solution, which means that the model can be applied in a spreadsheet.  Consequently, a more quantitative researcher could use the model as a starting point for their own model and tailor it to their study system.  For each new organism, empirically-measured parameter values can be provided as inputs and the corresponding minimum breeding population size computed using the analytical solution~\eqref{eq:NEWz1star}.  There is thus the potential for our approach to be broadly applied, helping to determine where many different populations are at risk.}
By developing region-specific strategies and prioritizing conservation while populations remain viable, we can prevent further losses and sustain these species for future generations.

\section*{Acknowledgements}
The modelling component and interpretation thereof was funded by the Natural Sciences and Engineering Research Council of Canada through an Undergraduate Summer Research Award (NDM) and Discovery grants (RGPIN-2016-05277 \& RGPIN-2022-03589, RCT), with support from the Columbia Basin Trust and Regional District of Central Kootenay Local Conservation Fund (MHM, WPM) through the Valhalla Wilderness Society. Additional funding for field research and data analysis was also provided by the Fish and Wildlife Compensation Program-Columbia Region and several private foundations.  This work was also supported by the UBC Okanagan Institute for Biodiversity, Resilience, and Ecosystem Services.  Numerous volunteer Toad Ambassadors assisted research biologists in collecting field data and removed thousands of toads from the road out of harm's way.  We thank Elke Wind and Barb Beasley for their comments which greatly improved an earlier draft of the manuscript, and three anonymous reviewers for their extensive and valuable comments.

\section*{Authorship}
As with many multidisciplinary papers, several authors made important contributions to the manuscript, and these cannot be properly described simply by relying on the author order provided.  NDM developed the model, established the analytical methods, and wrote the first draft of the paper.  RCT did extensive writing in all subsequent drafts of the paper, included the mortality data analysis, and funded the work by NDM and SKW.  MHM, together with WPM, conducted the field research and provided all of the data presented.  MHM met regularly with RCT to discuss the modelling work, and edited the paper extensively.  SKW verified the model equations and corrected the paper accordingly, and contributed to the editing of the final draft.  JM performed the statistical analysis of adult toad mortality, and contributed the associated text.  WPM also provided valuable editorial feedback.  Absolutely no AI was used in the generation of this manuscript.

\section*{Competing Interests}
The authors declare that they have no competing interests.

\section*{Data Availability}
The data is posted on GitHub at {\it (address to be added once the paper is accepted)}.

\newrefcontext[sorting=nty]
\printbibliography

\begin{appendices}

\section{Beverton-Holt growth function for females}
\label{sec:BHgrowth}

The standard Beverton-Holt growth function \citep{kot:2001} for the per-capita reproduction of a population $N$ with intrinsic growth rate $r$ and carrying capacity $K$ is written
\beq
\mathcal{B}(N) = \frac{rK}{K+(r-1)N}.
\label{eq:BHstandard}
\eeq
and the difference equation for population growth is
\beq
N_{t+1} = \mathcal{B}(N_t)N_t.
\label{eq:BHdiff-standard}
\eeq
We wish to adapt this function to a population where we only track the females.  {\RR This simplification makes the mathematical modelling easier, and is standard in population models where there are sufficient males to inseminate the females, and density-dependent effects can be ignored \citep{hostetler:2013, mollet:2002, halley:1996, crowder:1994, barbosa:2020}, and is often used in the context of determining population persistence.}

First, we assume that the population at time $t$ is half male, half female.  So the total population is twice the female population.  Second, we assume that the offspring are half female and half male.  Let $F_t$ and $M_t$ represent the female and male portions of the population at time $t$.  Then $N_t=F_t+M_t$ (and $N_{t+1}=F_{t+1}+M_{t+1}$) and~\eqref{eq:BHdiff-standard} becomes
\beq
2F_{t+1} = \frac{rK}{K+(r-1)2F_t} 2F_t \Leftrightarrow F_{t+1} = \mathcal{B}(2F_t) F_t.
\eeq
For our toad model therefore, we take the per-female birth rate of females to be $B(z)=\mathcal{B}(2z)$ where $z$ is the adult female population.  Note that the different fonts are intentional: $\mathcal{B}$ is the per capita birth rate, while $B$ is the per female birth rate.

\section{Extirpation Threshold}
\label{sec:extirpation}

The highway-crossing mortality level at which the population becomes locally extinct is given mathematically by setting $z_1^* = 0$ in~\eqref{eq:NEWz1star} and solving for the critical mortality level $m^c_z$.  We obtain
\beq
m_z^c = \frac{1}{2a}\left( b \pm \sqrt{b^2 - 4ac} \right)
\label{eq:survivorship-extinctThresh}
\eeq
where
\beqsub
\beqa
a & = & \zeta_g \left( Qr\delta\zeta_x - \alpha(\delta+\mu_y-\mu_z)e^{-\mu_zL} \right) \\ 
b & = & (\zeta_x+\zeta_g) - \alpha (\delta+\mu_y-\mu_z) e^{-\mu_zL} \\
c & = & Qr\delta - (\delta+\mu_y-\mu_z)
\eeqa
\eeqsub
and $Q$ is given in equation~\eqref{eq:Q}.  Note that $m_z^c$ is
independent of $K$, indicating that the extirpation threshold is independent of the carrying capacity.

\begin{figure}
    \centering
    \includegraphics[width=\textwidth]{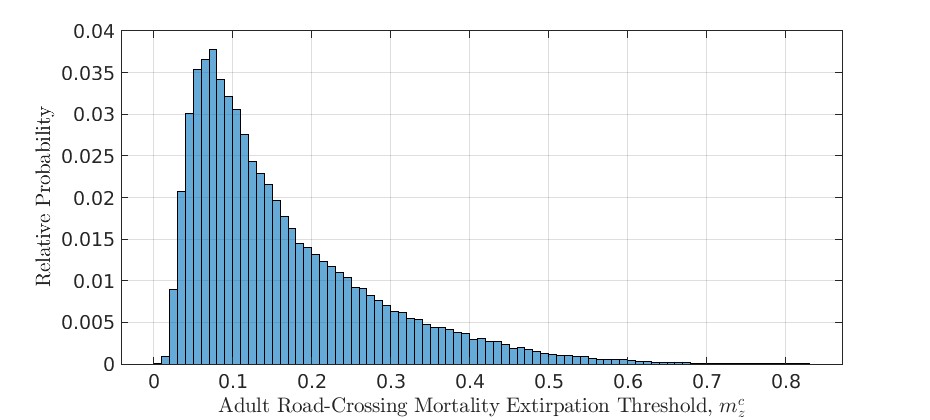}
    \caption{Histogram of the extirpation threshold for adult (non-gravid) highway-crossing mortality $m_z^c$ ranging from 0\%--85\% (0--0.85), taken from 100,000 repeated samples from the parameter space, with parameters distributed uniformly over the ranges defined in Table~\ref{tab:SymbolsTable}. 
    The extinction threshold mean is $m_z^c=0.16$ and median $m_z^c=0.13$.  The distribution maximum occurs at $0.07<m_z^c<0.08$.}
    \label{fig:extinctionThresh}
\end{figure}

Using Monte-Carlo sampling across the plausible range of parameter values (Table~\ref{tab:SymbolsTable}, we can compute the critical level of highway-crossing mortality for each sample.  Our results are shown in Figure~\ref{fig:extinctionThresh}.  We obtain a highly skewed distribution, with its maximum at $0.07<m_z^c<0.08$ and median at $m_z^c=0.13$.  So extirpation is more likely to occur sooner rather than later, i.e., for highway-crossing mortality levels on the order of 5-25\% rather than at higher levels.  

Note that while mathematical extinction~\eqref{eq:survivorship-extinctThresh} does not depend on lake carrying capacity, $K$, extinction is likely to occur with probability 1 for populations that are sufficiently small.  So, while our model predicts a population that is above zero for $m_z<m_z^c$, the real system certainly will be extirpated for highway-crossing mortality levels less than $m_z^c$. We can calculate a near extinction threshold by setting $z_1^*=V$ in~\eqref{eq:NEWz1star}, where $V$ can be set to, e.g., the value at which the population is considered endangered, or the value at which the population is considered threatened.  The resulting equation is cubic in $m_z$ and its solution gives us a mortality threshold value that does depend on $K$.

\end{appendices}

\makeatletter
\renewcommand \thesection{S\@arabic\c@section}
\renewcommand \thetable{S\@arabic\c@table}
\renewcommand \thefigure{S\@arabic\c@figure}
\renewcommand \theequation{S\@arabic\c@equation}
\makeatother

\newpage

\setcounter{section}{0}



\begin{center}
  {\Large{\bf{Supplement to: \\ Modelling road mortality risks to persistence to a Western Toad ({\it Anaxyrus boreas}) population in British Columbia}}} \\
\bigskip
{\large{Marguerite H. Mahr,  Noah D. Marshall, Jessa Marley, Sarah K. Wyse, Wayne P. McCrory, \& Rebecca C. Tyson}}
\end{center}

\section{Parameter Values}
\label{sec:paramVals}

For the purpose of analysis, we develop a range of values for each parameter in which the true value likely lies. Due to the paucity of relevant data, we make the parsimonious assumption of a uniform distribution for each parameter. To specify the distribution, we must define a minimum and a maximum.  If data is available, we select the mean of the data as the default value.  Otherwise, we select a value near the midpoint of the parameter range as the default value.

\bize
{\RR
\item ${\bm{K}}$: The carrying capacity $K$ is a key parameter in ecology \citep{zhang:2021}.  There are a number of definitions for carrying capacity \citep{chapman:2018}; here we use it to represent the well-established concept of the maximum population the environment can support. This definition is consistent with $K$ in the Beverton-Holt model.  Notwithstanding the clear mathematical definition, the carrying capacity of an environment is notoriously difficult to determine \citep{mcleod:1997}.  Spatial heterogeneity with dispersal \citep{zhang:2021} and temporal variability \citep{mcleod:1997} can both strongly influence the maximum population size.  Furthermore, recent work indicates that even in controlled experimental conditions, the carrying capacity cannot be predicted a priori \citep{subach:2023}.  We therefore provide our best estimate of the carrying capacity, given the information available, but allow $K$ to vary substantially from that value.  From the number of mating pairs observed during the field study (10-20 per night), the number of eggs per female (approximately 12,000, of which approximately 6,000 are female), and the length of the breeding period (approximately 30 nights) we arrive at 1.8--3.6 million eggs laid.  We thus use the following values for $K$:}
\begin{table}[H]
\centering
\begin{tabular}{@{}lll@{}}
\toprule
Minimum: 1.0 M & Default: 2.0 M & Maximum: 4.0 M \\ \bottomrule
\end{tabular}
\end{table}

\item ${\bm{m_z}}$: The parameter $m_z$ is the likelihood that the highway crossing will result in mortality for each adult non-gravid female migrating toad.  Since it was not always possible to determine whether a given female toad was gravid or not, we use the adult male data from non-COVID years to determine $m_z$.  The data are shown in Table~\ref{tbl:adult-male-mortality}.  {\RR Note that mortality rates were determined only using data gathered during the night-time observation periods, when both successful (live) and unsuccessful (dead) crossings could be counted.}

As there is considerable pressure to allow development in the area, traffic volumes could increase substantially. We therefore investigate population persistence under the full theoretical range of mortality ratios (from 0 to 1). 
\begin{table}[H]
\centering
\begin{tabular}{l|rrrr}
Year & live (\#) & dead (\#) & mortality (\%) & survivorship (\%)\\ \hline
2015 & 62 & 7 & 10\% & 90\% \\
2016 & 50 & 26 & 34\% & 66\% \\
2017 & 222 & 18 & 8\% & 92\% \\
2018 & 301 & 23 & 7\% & 93\% \\
2019 &  283 & 42 & 13\% & 87\% \\ \hline
2020 & 138 & 1 & 0.7\% & 99\% \\
2021 & 93 & 5 & 5\% & 95\% \\ \hline
\end{tabular}
\caption{Male adult toad highway-crossing data gathered {\RR during evening observation periods} at the Fish and Bear Lakes study area over the months April-July, before and during the COVID-19 pandemic in 2020-2021.} \label{tbl:adult-male-mortality}
\end{table}

The observation methods changed after 2016, and so we use the 2017-2019 data to compute the default mortality, and use the full data set to set the range.  We thus obtain the following minimum, default, and maximum values for $m_z$ under current traffic regimes, and interpret our results in terms of these values.
\begin{table}[H]
\centering
\begin{tabular}{@{}lll@{}}
\toprule
Minimum: 0.08 & Default: 0.1 & Maximum: 0.5 \\ \bottomrule
\end{tabular}
\end{table}

\item ${\bm{\zeta_x}}$: The $\zeta_x$ parameter represents the multiplicative increase in highway-crossing mortality for the toadlets compared to adults.  This parameter must be estimated as no data exist.  Toadlets tend to cross during the day, and later in the summer when tourist traffic is high.  Furthermore, the toadlets rarely cross alone.  Instead, they congregate on the lake side of the highway, until an unknown trigger sets off a mass migration of all of the toadlets crossing the highway at once.  Thus, one vehicle passing at that moment will kill a large number of toadlets.  We thus assume that toadlet crossings are more treacherous than adult crossings, so $\zeta_x>1$.  To compensate for the lack of data, we choose a broad range of values. 
\begin{table}[H]
\centering
\begin{tabular}{@{}lll@{}}
\toprule
Minimum: 1 & Default: 1.3 & Maximum: 2 \\ \bottomrule
\end{tabular}
\end{table}

\item ${\bm{\zeta_g}}$: The $\zeta_g$ parameter represents the multiplicative increase in highway-crossing mortality for adult gravid females compared to non-gravid adults.  We can estimate this value using the data gathered at Fish and Bear Lakes.  
Some adult females are seen crossing the highway in August; these are likely gravid females who are pre-emptively crossing before the winter in an effort to avoid the pressures of the spring migration.  As a first order approximation, we thus assume that the adult female mortality observed in August corresponds to the adult gravid female mortality.  Furthermore, assuming that adult non-gravid female mortality is the same as adult male mortality, the adult male mortality observed in August can be used as an estimate for adult non-gravid female mortality.  The data are shown in Table~\ref{tbl:Aug-mortality}.  Taking all of the data across the years 2016-2019, we observe that mortality for gravid females is nearly double that for males, i.e., $\zeta_g\approx2$.
\begin{table}[H]
\centering
\caption{Data on live and dead toads observed at the highway along Fish and Bear Lakes {\RR in August}. F refers to female counts, M to male counts.  The quantity ``mort" is the mortality calculated as the total number of dead F or M toads divided by the total number of live and dead F or M toads.}
\begin{tabular}{l|llll|ll}
\toprule
year & \# F live & \# M live & \# F dead & \# M dead & F mort. & M mort. \\
\hline
2016 & 25 & 14 & 14 & 8 \\
2017 & 32 & 10 & 42 & 2 \\
2018 & 10 & 3 & 11 & 5 \\
2019 & 14 & 7 & 9 & 2 \\ 
\hline
Totals & 81 & 34 & 76 & 17 & 30\% & 18\% \\
\bottomrule
\end{tabular}
\label{tbl:Aug-mortality}
\end{table}
\begin{table}[H]
\centering
\begin{tabular}{@{}lll@{}}
\toprule
Minimum: 1 & Default: 2 & Maximum: 3 \\ \bottomrule
\end{tabular}
\end{table}

\item ${\bm{\alpha}}$: \citet{olsonReturnBreeding} found 5.3\% of female toads returned to breed. In another study, however, approximately 8.5\% of female toads returned to the breeding site \citep{BreedingFrequencyOregon} . The Committee on the Status of Engangered Wildlife in Canada finds that only 5\% of females breed more than once \citep{cosewic:2012}, though this value appears to be elevation-dependent \citep{BreedingFrequencyOregon}.  Indeed, at one low altitude lake (1368m), the highest rate of return measured was 40 toads out of 158, or 24\% \citep{BreedingFrequencyOregon}.  With three lower values and one much higher one, we take the lowest and highest values as the enpoints of the plausible range, and select a default value toward the lower end.
\begin{table}[H]
\centering
\begin{tabular}{@{}lll@{}}
\toprule
Minimum: 0.053 & Default: 0.09 & Maximum: 0.24 \\ \bottomrule
\end{tabular}
\end{table}

\item ${\bm{r}}$: Clutch sizes measured in the literature are highly variable.  The BC Management Plan for Western toad populations states that clutch sizes range from 5,000-15,000 eggs \citep{MPforWT:2014}, while the provincial factsheet states that females lay clutches of up to 12,000 eggs \citep{zevit:2010}.  Among the published studies we found, one reports over 20,000 eggs in a single clutch \citep{maxell:2002}, while the remainder all report average clutch sizes below 12,000, with the majority falling in the range of 5,800-8,200 \citep{maxell:2002, biek:2002}. The number of observations in each of these studies varies considerably, from $n=4$ to $n=38$.  The study reporting over 20,000 eggs in one clutch has a sample size of $n=1$.  Taking the average clutch size from each of the studies reported in \citet{maxell:2002} and \citet{biek:2002}, and computing the overall average weighted by the sample size in each study, we arrive at an average clutch size of 6,839 eggs, or 3,420 female eggs (assuming a 1:1 ratio of males to females).  We found, however, that we needed a larger value in our simulations in order for the default parameter values to result in a persistent population above the threatened threshold.  We therefore take 12,000 eggs (or 6,000 female eggs) as the default clutch size.
%
\begin{table}[H]
\centering
\begin{tabular}{@{}lll@{}}
\toprule
Minimum: 3,000 & Default:  6,000 & Maximum: 7,000 \\ \bottomrule
\end{tabular}
\end{table}

\item ${\bm{\delta:}}$ This parameter gives the per day rate at which juveniles mature into adults. Juvenile females mature into adults after approximately 4-6 years \citep{govManPlan, BreedingFrequencyOregon}. By bounding $\delta$ such that 95\% of toads mature in 4 to 6 years, we establish a minimum and a maximum.
%
\begin{table}[H]
\centering
\begin{tabular}{@{}lll@{}}
\toprule
Minimum: 0.0014 & Default: 0.0017 & Maximum: 0.0021 \\ \bottomrule
\end{tabular}
\end{table}

\eize

\subsection{Mortality Rates}

As a toad matures, each stage has a lower mortality rate than the previous stage. Thus, $$\mu_z \leq \mu_y \leq \mu_x.$$
We estimate these rates by solving the equation $\frac{dP}{dt} = -\mu_PP$ for each population $P$.


\bize
\item $\bm{\mu_x}$:
Amphibians face high mortality during the first few months of their lives. The majority of eggs born will die. The percentage of eggs that survive until the summer is unknown. Expert opinion suggests that perhaps no less than 5\% and no more than 15\% survive the roughly 90 day maturation period. We take the midpoint as the default value. 
\begin{table}[H]
\centering
\begin{tabular}{@{}lll@{}}
\toprule
Minimum: 0.0211 & Default: 0.01 & Maximum: 0.0333 \\ \bottomrule
\end{tabular}
\end{table}

\item $\bm{\mu_y}$:
It is not possible to directly estimate the juvenile mortality rate. We bound $\mu_y$ above by the lowest value of $\mu_x$ and below by the highest value of $\mu_z$. 
\begin{table}[H]
\centering
\begin{tabular}{@{}lll@{}}
\toprule
Minimum: 0.0041  & Default: 0.0126 & Maximum: 0.0211\\ \bottomrule
\end{tabular}
\end{table}

\item $\bm{\mu_z}$:
Few estimates for the potential lifespan of western toads exist. \citet{cosewic:2012} suggests that the maximum lifespan of female western toads is approximately 9 years. As toads spend 4-7 years as juveniles, they spend 2-5 years as breeding adults.  We obtain a maximum mortality rate by assuming that 5\% of toads survive for 2 years (720 days), and a minimum mortality rate by assuming that 5\% of toads survive for 5 years (1800 days).  We take the midpoint of these two extremes as the default value.
\begin{table}[H]
\centering
\begin{tabular}{@{}lll@{}}
\toprule
Minimum: 0.0016 & Default: 0.00285 & Maximum: 0.0041 \\ \bottomrule
\end{tabular}
\end{table}

\eize

\section{Solution to Model Equations}
\label{sec:modelSol}

The model can be solved by considering this system as two systems of linear ODEs wherein the initial conditions for each system is given by the preceding impulse values.

\subsection{Egg-to-Toadlet Stage}
We first solve for the population during the egg-to-toadlet stage, i.e., when $t_1 + nL \leq t < t_2 + nL$.  We use the first of~\eqref{eq:breeding-crossing} to determine the initial condition for the first of~\eqref{eq:toadlet}.  Specifically, we take
\beq
x(t_1+nL) = B\left(\frac{(1-m_z)z(t^-)}{\zeta_g}\right) \left(\frac{(1-m_z)z(t^-)}{\zeta_g}\right)
\label{eq:toadlet-x-IC}
\eeq
Solving the ODE for $x(t)$, we obtain
\beq
x(t) =B\left(\frac{(1-m_z)z(t^-)}{\zeta_g}\right) \left(\frac{(1-m_z)z(t^-)}{\zeta_g}\right) e^{-\mu_x (t-(t_1 + nL))}
\label{eq:toadlet-x-soln}
\eeq
We can solve the second of~\eqref{eq:toadlet} directly, which yields
\beq 
y(t) = y(t_1 + nL) e^{-(\delta + \mu_y)(t - (t_1 + nL))}
\label{eq:toadlet-y-soln}
\eeq
Finally, we can solve the third of~\eqref{eq:toadlet} using an integrating factor, equation~\eqref{eq:toadlet-y-soln}, and the initial condition given by
\beq
z(t_1 + nL) = \frac{\alpha (1-\zeta_g m_z)(1-m_z) z((t_1 + nL)^-)}{\zeta_g}.
\label{eq:toadlet-z-IC}
\eeq
We obtain the solution
\beq
z(t) = - \, \frac{\delta \, y(t_1 + nL)}{\delta + \mu_y - \mu_z}e^{-(\delta + \mu_y)(t-(t_1 + nL))}
+ \left(z(t_1 + nL) + \frac{\delta \, y(t_1 + nL)}{\delta + \mu_y - \mu_z}\right)e^{-\mu_z(t-(t_1 + nL))} 
\label{eq:toadlet-z-soln}
\eeq
Together,~\eqref{eq:toadlet-x-soln}, \eqref{eq:toadlet-y-soln}, and~\eqref{eq:toadlet-z-soln} are the solutions for the summer period between the spring breeding and fall migration pulses.  

\subsection{The Fall and Winter Stage}

We obtain the solutions for the fall and winter stage in the same manner as we used above.  The time period for this stage is $t_2 + nL < t < t_1 + (n+1)L$. Clearly,
\beq
x(t) = 0, 
\label{eq:winter-x-soln}
\eeq
as the eggs have all matured into juveniles, and no new eggs are laid until the following spring.
The ODE for the juvenile population is the same as before (i.e., the second of~\eqref{eq:toadlet} is the same as the second of~\eqref{eq:winter}).  So the solution is
\beq 
y(t) = y(t_2 + nL) e^{-(\delta + \mu_y)(t - (t_2 + nL))}.
\label{eq:winter-y-soln}
\eeq
The equations for $z(t)$ (the third of~\eqref{eq:toadlet} and~\eqref{eq:winter}) are also the same, and so we can directly write
\beq
z(t) = -\frac{\delta \, y(t_2 + nL)}{\delta + \mu_y - \mu_z}e^{-(\delta + \mu_y)(t-(t_2 + nL))} + \left(z(t_2 + nL) + \frac{\delta \, y(t_2 + nL)}{\delta + \mu_y - \mu_z}\right)e^{-\mu_z(t-(t_2 + nL))}.
\label{eq:winter-z-soln}
\eeq
Together, equations~\eqref{eq:winter-x-soln}, \eqref{eq:winter-y-soln}, and~\eqref{eq:winter-z-soln} are the solutions for the population between the fall migration pulse and the spring breeding pulse.

\subsection{Full Solution}
Combining the ODE solutions with the discrete changes in population during the migration and breeding pulses we arrive at the full solution to~\eqref{eq:model}.  The solution is given by
\beqsub
\label{eq:NEWmodel_sol}
\begin{align}
\phantom{{\text{Breeding Migration }}}
\mathllap{\text{\bf Breeding Migration }} & {\text{\bf Impulse: }} t=t_1+nL, \nonumber \\
& \begin{aligned}
\mathllap{x(t)} & = B\left( (1-\zeta_g m_z) z(t^-) \right)\, (1-\zeta_g m_z)\, z(t^-), \\
\mathllap{y(t)} & = y(t^-), \\
\mathllap{z(t)} & = \alpha\, (1-\zeta_g m_z)\, (1-m_z)\, z(t^-)
\end{aligned}
\label{eq:NEWmodel_sol_BMI} \\
\nonumber \\
\mathllap{\text{\bf Egg to Toadlet Sta}} & {\text{\bf ge: }} t_1 + nL < t < t_2 + nL, \nonumber \\
& \begin{aligned}
\mathllap{x(t)} & = B\left( (1-\zeta_g m_z) z(t^-) \right) (1-\zeta_g m_z)z(t^-) e^{-\mu_x (t-(t_1 + nL))}, \\
\mathllap{y(t)} & = y(t_1 + nL) e^{-(\delta + \mu_y)(t - (t_1 + nL))}, \\
\mathllap{z(t)} & = -\frac{\delta \, y(t_1 + nL)}{\delta + \mu_y - \mu_z}  e^{-(\delta + \mu_y)(t-(t_1 + nL))} \\
& \qquad\qquad\qquad + \left( z(t_1 + nL) + \frac{\delta \, y(t_1 + nL)}{\delta + \mu_y - \mu_z}\right)e^{-\mu_z(t-(t_1 + nL))},
\end{aligned}
\label{eq:NEWmodel_sol_ET} \\
\nonumber \\
\mathllap{\text{\bf Toadlet Migration I}} & {\text{\bf mpulse: }} t = t_2 + nL, \nonumber \\
& \begin{aligned}
\mathllap{x(t)} & = 0, \\
\mathllap{y(t)} & = (1-\zeta_x m_z) \, x(t^-) + y(t^-), \\
\mathllap{z(t)} & = z(t^-), 
\end{aligned}
\label{eq:NEWmodel_sol_TMI} \\
\nonumber \\
\mathllap{\text{\bf Fall and Winter St}} & {\text{\bf age: }} t_2 + nL < t < t_1 + (n+1)L, \nonumber \\
& \begin{aligned}
\mathllap{x(t)} & = 0, \\
\mathllap{y(t)} & = y(t_2 + nL) e^{-(\delta + \mu_y)(t - (t_2 + nL))}, \\
\mathllap{z(t)} & =-\frac{\delta \, y(t_2 + nL) }{\delta + \mu_y - \mu_z} e^{-(\delta + \mu_y)(t-(t_2 + nL))} \\
& \qquad\qquad\qquad + \left ( z(t_2 + nL) + \frac{\delta \, y(t_2 + nL)}{\delta + \mu_y - \mu_z}\right)e^{-\mu_z(t-(t_2 + nL))}.
\end{aligned}
\label{eq:NEWmodel_sol_FW}
\end{align}
\eeqsub
 Note that, in the expressions above, we consider $\vec{x}(t_i+nL)$ to be the vector $(x(t),y(t),z(t))^T$ evaluated at the {\it{end}} of the impulse at the given time $t=t_i+nL$.

\section{Derivation of the periodic steady state solution}
\label{sec:NEWPerSteadyStatesDerivation}

In this section, we derive the periodic steady state of the model at the impulse times, $t_1$, $t_2$. First, we reformulate the model at times $t_1$, $t_2$ as a system of recursive sequences. Then we solve for the steady states.

Define the sequences
\beqsub
\label{eq:xseq}
\begin{align}
& x_1^n = x(t_1 + nL), n \in \mathbb{N} \cup \{0\} \\ 
& x_2^n = x(t_2 + nL), n \in \mathbb{N} \cup \{0\}
\end{align}
\eeqsub
Similarly define $y_{t_1}^n$, $y_{t_2}^n$,$z_{t_1}^n$,$z_{t_2}^n$. The limits of these sequences, when they exist, are periodic steady states of the system. They are reoccurring population values at specific times of the year, i.e., at the migrations. Denote the limit as $t \to \infty$ of the above sequences with a star (*) superscript.

At the periodic steady-state, the system~\eqref{eq:NEWmodel_sol} yields the solutions immediately following each impulse.  We obtain
\beqsub
\label{eq:NEWmodel_steadystate}
\begin{align}
\phantom{{\text{Breeding Migration }}}
\mathllap{\text{\bf Breeding Migration }} & {\text{\bf Impulse: }} t=t_1+nL, \nonumber \\
& \begin{aligned}
\mathllap{x_1^*} & = B\left( \frac{z_1^*}{\alpha(1-m_z)} \right)\, \frac{z_1^*}{\alpha(1-m_z)}, 
\\
\mathllap{y_1^*} & = y_2^* e^{-(\delta+\mu_y)(T+L)}, \\
\mathllap{z_1^*} & = - \, \frac{\delta \, y_2^*}{\delta+\mu_y-\mu_z} e^{-(\delta+\mu_y)(L-T)}
+
\left[ z_2^* - \frac{\delta \, y_2^*}{\delta+\mu_y-\mu_z} \right] e^{-\mu_z (L-T)}
\end{aligned}
\label{eq:NEWmodel_ss_BMI} \\
\nonumber \\
\mathllap{\text{\bf Toadlet Migration I}} & {\text{\bf mpulse: }} t = t_2 + nL, \nonumber \\
& \begin{aligned}
\mathllap{x_2^*} & = 0, \\
\mathllap{y_2^*} & = (1-\zeta_x m_z) \, x_1^* \, e^{-\mu_x T} + y_1^* \, e^{-(\delta+\mu_y)T}, \\
\mathllap{z_2^*} & = - \, \frac{\delta}{\delta+\mu_y-\mu_z} \, y_1^* \, e^{-(\delta+\mu_y)T} 
+
\left[ z_1^* - \frac{\delta}{\delta+\mu_y-\mu_z} \, y_1^* \right] e^{-\mu_z T}, 
\end{aligned}
\label{eq:NEWmodel_ss_TMI} 
\end{align}
\eeqsub
Solving~\eqref{eq:NEWmodel_steadystate} for $z_1^*$, we obtain~\eqref{eq:NEWz1star}, which we reproduce here, for ease of reference:
\beq
z_1^* = \frac{K\alpha(1-m_z)}{2(r-1)} 
\left(
\frac{Qr\delta}{\delta+\mu_y-\mu_z} \, 
\frac{(1-\zeta_x m_z)(1-\zeta_gm_z)}{1-\alpha(1-\zeta_gm_z) \, m_z \, e^{-\mu_zL}}
-1
\right)
\label{eq:NEWz1star-app}
\eeq
where $Q$, given by~\eqref{eq:Q}, is a factor involving only the death rates $\mu_i$, $i\in{x, y, z}$, the juvenile maturation rate $\delta$, and the time intervals $L$ and $T$.

To ensure that the constraints~\eqref{eq:zeta-constraint} are satisfied, we translate~\eqref{eq:NEWz1star-app} in terms of survivorship $s_z=1-m_z$ and factors reducing survivorship, $\eta_i s_z=1-\zeta_i m_z$ where $i\in{g,x}$ to obtain
\beq
z_{t_1}^* = \frac{K\alpha s_z}{2(r-1)} 
\left(\frac{Qr \delta}{\delta+\mu_y-\mu_z} \, 
\frac{\eta_x\eta_gs_z^2}{1-\alpha\eta_gs_z^2e^{-\mu_zL}} -  1
\right).
\label{eq:NEWz1star-app-survivorship}
\eeq
For our simulations, we use~\eqref{eq:NEWz1star-app-survivorship}, varying $s_z$, $\eta_g$, and $\eta_x$, and then present the results in terms of the more intuitive parameters $m_z$, $\zeta_g$, and $\zeta_x$.

\section{Generalized Linear Models}
\label{sec:regression}

\subsection{Data}
\label{app:modeldata}
Data were collected from 2016 to 2021 for the months of May through October at the latest. The number of dead toads found on the road during the collection period was recorded, along with the number of vehicles. If a live toad was found, it was counted and moved off the roadway. The length of the collection period varied per collection period, measured  in minutes. The length of the collection period varies between 15 minutes to 185 minutes.

The data was categorized into the start time categories of ``evening" or ``night" by treating any collection start time before 10:00pm as ``evening" and after 10:00pm as ``night". There was also a categorization for the years as an indicator for pre-COVID and during COVID, where pre-COVID is before 2020 and during COVID is 2020 and later.

We assume a Poisson distribution for all responding variables, count of vehicles, count of live toads, and count of dead toads, due to the discrete and non-negative nature of the count data.  For all fitted models, an offset is included to adjust for the amount of time available for data collection. We used a generalized linear model and are fitting a Poisson distribution, with year as a random effect. We considered a zero-inflated model, and found that all distributions did not meet the typical qualifications for this types of data.

\subsection{Vehicle Model Results}
\label{app:vehiclemodel}

In order to gain insight into the use of the road by vehicles, we modeled the number of vehicles seen in the collection period. The results show us when to expect increased traffic and give a better understanding of what may be influencing factors.

Testing several models, including the null model, the results are shown in Table~\ref{table:vehicles}.

\begin{longtable}{l|r|rrrr} 
\caption{Generalized linear model results for fitting a Poisson distribution to the number of vehicles on the road during the collection time by varying explanatory variables. We use AIC as the metric to evaluate model of best fit.} 
\label{table:vehicles}
\endfirsthead
\endhead
Model  & AIC & Covariates & Coeff. Est. & Std. Error & Sig. ($p$)\\ \hline
$\text{\# of vehicles}\sim \beta_0$   & 801.3 & Intercept & -2.8889 & 0.156 & $<2\times 10^{-16}$ \\ 
 $ + (1|\text{year})$ &  &  &  &  &  \\  \hline
$\text{\# of vehicles}\sim \beta_0 $ & 730.2 & Intercept & -3.20 & 0.25 & $<2\times 10^{-16}$ \\ 
 $ + \text{month} + (1|\text{year})$ &  & Month, Aug. & 0.80 & 0.24 & 0.00095 \\ 
  &  & Month, July & 0.64 & 0.24 & 0.0090 \\ 
 &  & Month, June & 0.064 & 0.25 & 0.80 \\ 
 &  & Month, May & -0.066 & 0.23 & 0.78 \\ 
 &  & Month, Sept. & 0.64 & 0.24 & 0.0084 \\ \hline
 $\text{\# of vehicles}\sim \beta_0 $ & 793.0 & Intercept & -3.36 & 0.12 & $<2\times 10^{-16}$ \\ 
  $+ \text{COVID} + (1|\text{year})$ &  & Pre-COVID & 0.72 & 0.14 & $2.03\times 10^{-7}$ \\ \hline 
  $\text{\# of vehicles}\sim \beta_0 $ & 787.8 & Intercept & -2.63 & 0.14 & $<2\times 10^{-16}$ \\ 
  $+ \text{Start time cat.} + (1|\text{year})$ &  & Start, night & -0.35 & 0.089 & $7.48\times 10^{-5}$ \\ \hline 
  $\text{\# of vehicles}\sim \beta_0 $ & 724.6 & Intercept & -3.52 & 0.25 & $<2\times 10^{-16}$ \\ 
  $ + \text{month}$ &  & Month, Aug. & 0.74 & 0.24 & 0.0023 \\ 
  $+ \text{COVID} + (1|\text{year})$ &  & Month, July & 0.59 & 0.24 & 0.015 \\ 
  &  & Month, June & 0.029 & 0.25 & 0.91 \\ 
  &  & Month, May & -0.11 & 0.24 & 0.65 \\
  &  & Month, Sept. & 0.57 & 0.25 & 0.020 \\
  &  & Pre-COVID & 0.54 & 0.14 & 0.00012 \\ \hline
  $\text{\# of vehicles}\sim \beta_0 $ & 722.8 & Intercept & -3.07 & 0.25 & $<2\times 10^{-16}$ \\ 
  $ + \text{month}$ &  & Month, Aug. & 0.87 & 0.24 & 0.00032 \\ 
  $+ \text{start time cat.} + (1|\text{year})$ &  & Month, July & 0.87 & 0.26 & 0.00065 \\ 
  &  & Month, June & 0.26 & 0.26 & 0.31 \\ 
  &  & Month, May & 0.057 & 0.24 & 0.81 \\ 
  &  & Month, Sept. & 0.58 & 0.24 & 0.017 \\
  &  & Start, night & -0.36 & 0.12 & 0.0021 \\ \hline
  $\text{\# of vehicles}\sim \beta_0 $ & 780.5 & Intercept & -3.047 & 0.13 & $<2\times 10^{-16}$ \\ 
  $ + \text{COVID}$ &  & Pre-COVID & 0.59 & 0.13 & $5.43\times10^{-6}$ \\ 
  $+ \text{start time cat.} + (1|\text{year})$ &  & Start, night & -0.34 & 0.087 & 0.00013 \\ \hline
$\text{\# of vehicles}\sim \beta_0 $ & 717.9 & Intercept & -3.34 & 0.25 & $<2\times 10^{-16}$ \\ 
  $ + \text{month}$ &  & Month, Aug. & 0.80 & 0.24 & 0.00094 \\ 
 $ + \text{COVID}$ &  & Month, July & 0.82 & 0.26 & 0.0014 \\ 
  $+ \text{start time cat.} + (1|\text{year})$ &  & Month, June & 0.22 & 0.26 & 0.39 \\ 
  &  & Month, May & 0.015 & 0.24 & 0.9t \\ 
  &  & Month, Sept. & 0.51 & 0.25 & 0.041 \\
  &  & Pre-COVID & 0.45 & 0.13 & 0.00058 \\
  &  & Start, night & -0.34 & 0.11 & 0.0031 \\ \hline
\end{longtable}

Using the best fitting model, we predict the number of vehicles expected on the road for each month, either in the morning or the evening and either pre- or during COVID. These results are shown in Table~\ref{table:vehicleprediction}.

\begin{table}[H]
\centering
\caption{The predicted number of vehicles per hour under the given conditions according to the model of best fit, which has month, COVID indicator, and start time category as the covariates.} \label{table:vehicleprediction}
\begin{tabular}{l|rrrr}
Month & Pre-COV, Evening & COV Evening & Pre-COV, Night & COV, Night\\ \hline
April & 3.35 & 2.14 & 2.38 & 1.52 \\
May & 3.40 & 2.17 & 2.42 & 1.54 \\
June & 4.17 & 2.66 & 2.97 & 1.90 \\
July & 7.57 & 4.83 & 5.39 & 3.44 \\
August &  7.46 & 4.76 & 5.31 & 3.39 \\ 
September & 4.28 & 2.73 & 3.05 & 1.95 \\ \hline
\end{tabular}
\end{table}

\subsection{Live Toad Model Results}
\label{app:livetoadmodel}

The live toad model looks at the relationship between the number of live toads on the road and the covariates. We follow the same process as in Section~\ref{app:vehiclemodel}. Tested models and fitted values are shown in Table~\ref{table:livetoads}.

\begin{longtable}{l|r|rrrr}
\caption{Generalized linear model results for fitting a Poisson distribution to the number of live toads on the road during the collection time by varying explanatory variables. We use AIC as the metric to evaluate model of best fit.} 
\label{table:livetoads}
\endfirsthead
\endhead
Model  & AIC & Covariates & Coeff. Est. & Std. Error & Sig. ($p$)\\ \hline
$\text{\# of live}\sim \beta_0$   & 1407.3 & Intercept & -2.054 & 0.12 & $<2\times 10^{-16}$ \\ 
 $ + (1|\text{year})$ &  &  &  &  &  \\  \hline
 
$\text{\# of live}\sim \beta_0 $ & 1194.8 & Intercept & -2.24 & 0.19 & $<2\times 10^{-16}$ \\ 
 $ + \text{month} + (1|\text{year})$ &  & Month, August & -0.18 & 0.17 & 0.27 \\ 
  &  & Month, July & -0.60 & 0.18 & 0.00081 \\ 
 &  & Month, June & 0.024 & 0.16 & 0.88 \\ 
 &  & Month, May & 0.56 & 0.15 & 0.00011 \\ 
 &  & Month, Sept. & -0.055 & 0.16 & 0.74 \\ \hline
 
 $\text{\# of live}\sim \beta_0 $ & 1398.7 & Intercept & -2.43 & 0.085 & $<2\times 10^{-16}$ \\ 
  $+ \text{COVID} + (1|\text{year})$ &  & Pre-COVID & 0.57 & 0.10 & $2.76\times 10^{-8}$ \\ \hline 

  $\text{\# of live}\sim \beta_0 $ & 1372.7 & Intercept & -1.85 & 0.15 & $<2\times 10^{-16}$ \\ 
  $+ \text{Vehicles} + (1|\text{year})$ &  & Vehicles & -0.039 & 0.0070 & $7.04\times 10^{-9}$ \\ \hline 

  $\text{\# of live}\sim \beta_0 $ & 1409.3 & Intercept & -2.058 & 0.13 & $<2\times 10^{-16}$ \\ 
  $+ \text{Start cat.} + (1|\text{year})$ &  & Start, night & 0.0050 & 0.061 & 0.94 \\ \hline 
  
  $\text{\# of live}\sim \beta_0 $ & 1179.8 & Intercept & -2.61 & 0.15 & $<2\times 10^{-16}$ \\ 
  $ + \text{month}$ &  & Month, Aug. & -0.24 & 0.16 & 0.14 \\ 
  $+ \text{COVID} + (1|\text{year})$ &  & Month, July & -0.66 & 0.18 & 0.00021 \\ 
  &  & Month, June & -0.0055 & 0.16 & 0.97 \\ 
  &  & Month, May & 0.52 & 0.15 & 0.00032 \\
  &  & Month, Sept. & -0.10 & 0.16 & 0.53 \\
  &  & Pre-COVID & 0.64 & 0.065 & $<2\times 10^{-16}$ \\ \hline

  $\text{\# of live}\sim \beta_0 $ & 1183.4 & Intercept & -2.16 & 0.20 & $<2\times 10^{-16}$ \\ 
  $ + \text{month}$ &  & Month, Aug. & -0.085 & 0.17 & 0.61 \\ 
  $+ \text{vehicles} + (1|\text{year})$ &  & Month, July & -0.54 & 0.18 & 0.0026 \\ 
  &  & Month, June & 0.039 & 0.16 & 0.80 \\ 
  &  & Month, May & 0.59 & 0.15 & $5.57\times10^{-5}$ \\
  &  & Month, Sept. & 0.023 & 0.16 & 0.89 \\
  &  & Vehicles & -0.024 & 0.0068 & 0.00037 \\ \hline

   $\text{\# of live}\sim \beta_0 $ & 1196.5 & Intercept & -2.22 & 0.19 & $<2\times 10^{-16}$ \\ 
  $ + \text{month}$ &  & Month, Aug. & -0.18 & 0.17 & 0.29 \\ 
  $+ \text{start cat.} + (1|\text{year})$ &  & Month, July & -0.57 & 0.18 & 0.0018 \\ 
  &  & Month, June & 0.044 & 0.16 & 0.79 \\ 
  &  & Month, May & 0.58 & 0.15 & $9.71\times10^{-5}$ \\
  &  & Month, Sept. & -0.066 & 0.16 & 0.69 \\
  &  & Start, night & -0.039 & 0.073 & 0.59 \\ \hline

  $\text{\# of live}\sim \beta_0 $ & 1364.5 & Intercept & -2.30 & 0.10 & $<2\times 10^{-16}$ \\ 
  $ + \text{COVID}$ &  & Pre-COVID & 0.68 & 0.13 & $6.63\times 10^{-8}$ \\ 
  $+ \text{vehicles} + (1|\text{year})$ &  & Vehicles & -0.039 & 0.0067 & $9.92\times 10^{-9}$ \\ \hline

   $\text{\# of live}\sim \beta_0 $ & 1400.7 & Intercept & -2.43 & 0.10 & $<2\times 10^{-16}$ \\ 
  $ + \text{COVID}$ &  & Pre-COVID & 0.58 & 0.11 & $4.9\times 10^{-8}$ \\ 
  $+ \text{Start cat.} + (1|\text{year})$ &  & Vehicles & 0.007 & 0.060 & 0.91 \\ \hline

  $\text{\# of live}\sim \beta_0 $ & 1170.1 & Intercept & -2.57 & 0.15 & $<2\times 10^{-16}$ \\ 
  $ + \text{month}$ &  & Month, Aug. & -0.16 & 0.17 & 0.35 \\ 
  $+ \text{COVID} $ &  & Month, July & -0.61 & 0.18 & 0.00081 \\ 
  $+ \text{vehicles} + (1|\text{year})$&  & Month, June & 0.0034 & 0.16 & 0.98 \\ 
  &  & Month, May & 0.54 & 0.15 & 0.00021 \\
  &  & Month, Sept. & -0.028 & 0.17 & 0.87 \\
  &  & Pre-COVID & 0.70 & 0.077 & $<2\times 10^{-16}$ \\
  &  & Vehicles & -0.022 & 0.0066 & 0.0011 \\ \hline

  $\text{\# of live}\sim \beta_0 $ & 1181.6 & Intercept & -2.59 & 0.15 & $<2\times 10^{-16}$ \\ 
  $ + \text{month}$ &  & Month, Aug. & -0.24 & 0.17 & 0.15 \\ 
  $+ \text{COVID} $ &  & Month, July & -0.64 & 0.18 & 0.00052 \\ 
  $+ \text{start cat.} + (1|\text{year})$&  & Month, June & 0.015 & 0.16 & 0.93 \\ 
  &  & Month, May & 0.54 & 0.15 & 0.00028 \\
  &  & Month, Sept. & -0.11 & 0.16 & 0.50 \\
  &  & Pre-COVID & 0.64 & 0.067 & $<2\times 10^{-16}$ \\
  &  & Start, night & -0.034 & 0.069 & 0.62 \\ \hline

  $\text{\# of live}\sim \beta_0 $ & 1182.3 & Intercept & -2.088 & 0.20 & $<2\times 10^{-16}$ \\ 
  $ + \text{month}$ &  & Month, Aug. & -0.051 & 0.17 & 0.76 \\ 
  $+ \text{start cat.} $ &  & Month, July & -0.45 & 0.19 & 0.016 \\ 
  $+ \text{vehicles} + (1|\text{year})$&  & Month, June & 0.11 & 0.16 & 0.50 \\ 
  &  & Month, May & 0.64 & 0.15 & $1.70\times 10^{-5}$ \\
  &  & Month, Sept. & -0.0014 & 0.17 & 0.99 \\
  &  & Start, night & -0.14 & 0.077 & $8.72\times 10^{-5}$ \\
  &  & Vehicles & -0.028 & 0.0070 & 0.079 \\ \hline

  $\text{\# of live}\sim \beta_0 $ & 1363.7 & Intercept & -2.19 & 0.12 & $<2\times 10^{-16}$ \\ 
  $ + \text{COVID}$ &  & Pre-COVID & 0.66 & 0.12 & $7.48\times10^{-8}$ \\ 
  $+ \text{start cat.} $ &  & Vehicles & -0.042 & 0.0070 & $2.77\times10^{-9}$ \\ 
  $+ \text{vehicles} + (1|\text{year})$&  & Start, night & -0.11 & 0.063 & 0.094 \\ \hline

  $\text{\# of live}\sim \beta_0 $ & 1169.2 & Intercept & -2.49 & 0.16 & $<2\times 10^{-16}$ \\ 
  $ + \text{month}$ &  & Month, Aug. & -0.13 & 0.17 & 0.45 \\ 
  $+ \text{start cat.} $ &  & Month, July & -0.52 & 0.19 & 0.0055 \\ 
  $ + \text{COVID}$ &  & Month, June & 0.074 & 0.16 & 0.65 \\ 
  $+ \text{vehicles} + (1|\text{year})$&  & Month, May & 0.59 & 0.15 & $7.3\times 10^{-5}$ \\
  &  & Month, Sept. & -0.047 & 0.17 & 0.77 \\
  &  & Start, night & -0.13 & 0.074 & 0.086 \\
  &  & Vehicles & -0.025 & 0.0069 & 0.00028 \\
  &  & Pre-COVID & 0.68 & 0.074 & $<2\times 10^{-16}$ \\ \hline
  
\end{longtable}

The results indicate that there may be a relationship between the number of toads on the road and the number of vehicles on the road, which is showing that more vehicles result in less toads on the road. This effect may be impacted by researchers ability to count toads when the roads are busier, or it could indicate that less toads use the road when it's busier, or another unknown factor. We also note that the impacts pre- and during COVID could be due to the impacts COVID had on the researcher's ability to gather data or other unknown outside factors

Similarly to the above section, we predict the number of live toads present for each month using the model of best fit. For these results, see Table~\ref{table:livetoadsprediction}.

\begin{longtable}{l|r|rrrr} 
Month & Pre-COV, Evening & COV Evening & Pre-COV, Night & COV, Night\\ \hline
April & 9.80 & 4.97 & 8.63 & 4.34  \\
May & 17.74 & 9.00 & 15.63 & 7.93  \\
June & 10.55 & 5.36 & 9.30 & 4.72  \\
July & 5.82 & 2.95 & 5.12 & 2.60 \\
August & 8.622  & 4.38 & 7.60 & 3.86 \\ 
September & 9.34 & 4.74 & 8.23 & 4.18 \\ \hline
\caption{The predicted number of live toads per hour under the given conditions according to the model of best fit, which has month, COVID indicator, and start time category as the covariates.} \label{table:livetoadsprediction}
\end{longtable}

\subsection{Dead Toad Model Results}

When we consider the possible impacts on toad mortality, such as the month itself and pre/during COVID, these factors don't have an obvious impact on the death of the toads aside from impacting the number of vehicles on the road and the number of live toads on the road. As a result, we focus on the models shown in Table~\ref{table:deadtoads} to compare to the null model. We do consider the start time category as it is possible that it may impact the number of toads dead due to visibility and drivers avoiding toads. 

\begin{longtable}{l|r|rrrr}
Model  & AIC & Covariates & Coefficient Est. & Std. Error & Sig. ($p$)\\ \hline
$\text{\# of dead}\sim \beta_0$   & 397.3 & Intercept & -4.20 & 0.42 & $<2\times 10^{-16}$ \\ 
 $ + (1|\text{year})$ &  &  &  &  &  \\  \hline

 $\text{\# of dead}\sim \beta_0 $ & 389.6 & Intercept & -4.56 & 0.42 & $<2\times 10^{-16}$ \\ 
  $+ \text{Live} $ &  & Num. Live & 0.011 & 0.0036 & 0.0019 \\ 
  $+ \text{Vehicles} + (1|\text{year})$ &  & Vehicles & 0.030 & 0.013 & 0.020 \\ \hline 

  $\text{\# of dead}\sim \beta_0 $ & 391.5 & Intercept & -4.51 & 0.44 & $<2\times 10^{-16}$ \\ 
  $+ \text{Live} $ &  & Num. Live & 0.011 & 0.0036 & 0.0023 \\ 
  $+ \text{Start cat.} $ &  & Start, night & -0.053 & 0.016 & 0.74 \\ 
  $+ \text{Vehicles} + (1|\text{year})$ &  & Vehicles & 0.030 & 0.013 & 0.028 \\\hline 
\caption{Generalized linear model results for fitting a Poisson distribution to the number of dead toads on the road during the collection time by varying explanatory variables. We use AIC as the metric to evaluate model of best fit.} \label{table:deadtoads}
\end{longtable}

From the best fitting model, with number of vehicles and number of live toads as the explanatory variables, we determine that for each additional vehicle on the road during a given time period (here time is unitless), there is a 3.1\% increase in the number of dead toads. For context, for each additional live toad on the road, there is an increase of 1.1\% in the number of dead toads resulting.

\end{document}